\def\squeeze{0}
\def\confversion{0}
\documentclass{vldb}

\usepackage{balance}
\usepackage{nopageno}

\usepackage[T1]{fontenc}
\usepackage{url}
\usepackage{hyperref}
\usepackage{color}
\usepackage{xspace}
\usepackage{graphicx}
\usepackage{mathptmx} 
\usepackage{balance}
\usepackage{enumitem}
\usepackage{amsmath,amssymb}
\setlist{nosep}

\newtheorem{theorem}{Theorem}

\DeclareGraphicsExtensions{.pdf,.eps}

\def\hv{\usefont{OT1}{phv}{m}{n}\selectfont}

\def\t#1{\textit{#1}}

\def\helvsize{\small}

\def\hv{\usefont{OT1}{phv}{m}{n}\selectfont\helvsize}

\makeatletter
\def\url@newstyle{\def\UrlFont{\hv\scriptsize}}
\makeatother
\def\eg{e.g.}

\if\squeeze 1
  \newcommand\xparagraph[1]{\vspace{1mm}\noindent \textbf{#1}}

\else
  \newcommand\xparagraph[1]{\vspace{2mm}\noindent \textbf{#1}}

\fi

\def\sysname{Hill\-view\xspace}

\def\P{\mathcal{P}}

\newenvironment{Proof}{\medbreak
\noindent {\bf Proof:~}}{\unskip\nobreak\hfill\hskip 2em \qed\par\medbreak}

\newcommand{\eat}[1]{}

\newcommand{\rgta}{\ensuremath{\rightarrow}}

\newcommand{\crt}{\t{summarize}}
\newcommand{\agg}{\t{merge}}

\newcommand{\ignore}[1]{}

\def\colorful{3}

\ifnum\colorful=1

\fi

\newcommand{\F}{\mathcal F}

\newcommand{\D}{\mathcal D}

\newcounter{this-list}

\vldbTitle{A Sample Proceedings of the VLDB Endowment Paper in LaTeX Format}
\vldbAuthors{Mihai Budiu, Parikshit Gopalan, Lalith Suresh, Udi
Wieder, Han Kruiger, and Marcos K. Aguilera}
\vldbDOI{https://doi.org/10.14778/3342263.3342279}
\vldbVolume{12}
\vldbNumber{11}
\vldbYear{2019}

\begin{document}

\title{\sysname: A trillion-cell spreadsheet for big data}
\numberofauthors{6} 
\author{
  \alignauthor
  Mihai Budiu \\
  \email{mbudiu@vmware.com} \\
  \affaddr{VMware Research}
  \alignauthor
  Parikshit Gopalan \\
  \email{pgopalan@vmware.com} \\
  \affaddr{VMware Research}
  \alignauthor
  Lalith Suresh \\
  \email{lsuresh@vmware.com} \\
  \affaddr{VMware Research}
\and
  \alignauthor
  Udi Wieder \\
  \email{uwieder@vmware.com} \\
  \affaddr{VMware Research} \\
  \alignauthor
  Han Kruiger \\
  \affaddr{University of Utrecht} \\
 \alignauthor
  Marcos K. Aguilera \\
  \email{maguilera@vmware.com} \\
  \affaddr{VMware Research} \\
}

\maketitle

\begin{abstract}

  \sysname is a distributed spreadsheet for browsing
  very large datasets that cannot be handled by a single machine.
As a spreadsheet, \sysname provides a high degree of interactivity that
  permits data analysts to explore information quickly along many
  dimensions while switching visualizations on a whim.
To provide the required responsiveness, \sysname introduces
  visualization sketches, or {\em vizketches}, as a simple idea
  to produce compact data visualizations.
Vizketches combine
  algorithmic techniques for data summarization with
  computer graphics principles for efficient rendering.
While simple, vizketches are effective at scaling the spreadsheet
  by parallelizing computation, reducing communication, providing
  progressive visualizations, and offering precise accuracy
  guarantees.
Using \sysname running on eight servers, we can navigate and visualize
  datasets of tens of billions of rows and trillions of cells, much beyond
  the published capabilities of competing systems.

\end{abstract}

\section{Introduction}

Enterprise systems store valuable data about
  their business.
For example, retailers store data about purchased items;
  credit card companies, about transactions;
  search engines, about queries; and
airlines, about flights and passengers.
To understand this data, companies hire data analysts whose job
  is to extract deep business insights.
To do that, analysts like to use spreadsheets such as Excel, Tableau, or PowerBI,
  which serve to
  explore the data {\em interactively}, by plotting charts, zooming in,
  switching charts, inspecting raw data, and repeating.
Rapid interaction distinguishes spreadsheets from other solutions, such as
  analytics platforms and batch-based systems.
Interaction is desirable, because the analyst does not know initially where to look, so she must
  explore data quickly along many dimensions and change visualizations on a whim.

Unfortunately, enterprise data is growing dramatically, and
  current spreadsheets do not work with big data, because
  they are limited in capacity or interactivity.
Centralized spreadsheets such as Excel can handle only millions of rows.
More advanced tools such as Tableau can scale to larger data sets by
  connecting a visualization front-end
  to a general-purpose analytics engine in the back-end.
Because the engine is general-purpose,
  this approach is either slow for a big data spreadsheet
  or complex to use as it requires users to carefully choose 
  queries that the system is able to execute quickly.
For example, Tableau can use Amazon Redshift as the analytics
  back-end but users must understand lengthy documentation
  to navigate around bad query types that are too 
  slow to execute~\cite{christopher-tableau19}.

We propose {\em \sysname}, a distributed spreadsheet for big data.
\sysname can navigate and visualize hundreds of columns and
  tens of billions of rows, totaling a trillion cells,
  far beyond the capability of the best interactive tools today.
\sysname uses a distributed system with worker servers that provide storage
  and computation.
It achieves massive data scalability with just a few servers (e.g.,
with eight commodity servers it supports a trillion spreadsheet cells).

The main challenge facing \sysname is to
  provide near real-time performance despite having to compute
  over big data.

To address this challenge, \sysname invokes a common idea in data\-base
  design: specialize the engine~\cite{endofarchitectural}.
Rather than using a general-purpose analytics engine,
  \sysname introduces a new engine specialized
  to render the tabular views and charts of a spreadsheet.
The main technical novelty of the paper is how to accomplish this specialization:
  we introduce the notion of  \emph{visualization sketches} or
  simply {\em vizketches}, and we propose a
  new distributed engine to render visualizations quickly using vizketches.
  
Vizketches combine
  ideas from the algorithms
  and computer graphics communities.  
In the algorithms community, {\em mergeable summaries}~\cite{Agarwal2012}
  are approximate computations that compute results
  over disjoint subsets of the data,
  that can then be merged to obtain the final result.
Mergeable summaries are useful to distribute the computation efficiently with
  fine control over the accuracy and resolution of the result.
A vizketch combines mergeable summaries with a basic principle in
  computer graphics rendering:
  {\em compute only what you can display}.
A vizketch, thus, adjusts its accuracy and resolution to match the display resolution and
  compute only what can be visually discerned.
For example, a vizketch for producing histograms
  limits the number of bars to ${\approx}$100
  and computes the height of each bar only to the nearest pixel; these choices
  reduce the network communication and enable computation over big data.
If the user zooms in on the histogram, the vizketch adapts to the new
  visualization to adjust the histogram buckets and
  enhance the accuracy of the bars while avoiding
  the computation of bars that are no longer visible.
  
Vizketches play a crucial role in \sysname.
They parallelize the computation, reduce communication bandwidth,
  enhance computation efficiency,
 permit a progressive visualization of
 results, provide a precise accuracy guarantee,
  and ensure scalability (\S\ref{sec:benefits}).
These benefits are key for a spreadsheet to be able to browse big data at interactive speeds.
Furthermore, vizketches \emph{can always} be computed efficiently.
This feature differentiates \sysname from traditional visualization solutions,
  which let users specify broad declarative queries without exposing their
  performance to the user, which is problematic for efficiency or usability.
That leads to an important question about \sysname: are the queries supported
   by vizketches rich enough to implement a fully functional
 spreadsheet?
A contribution of this paper is to answer this question positively.

\begin{figure}
  \begin{center}
  	\includegraphics[width=2.5in]{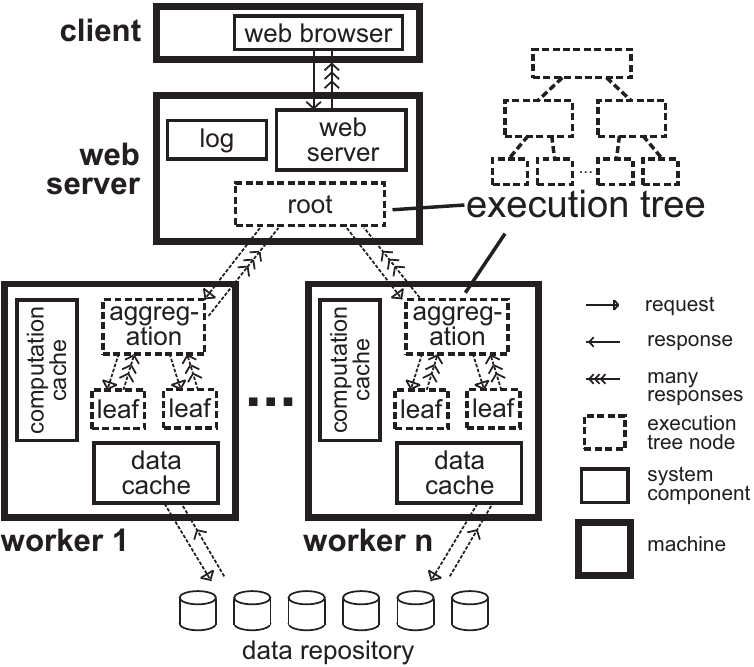}
  \end{center}
 \caption{\sysname is a spreadsheet for browsing big data. It introduces a novel
    database engine based on vizketches to distribute, parallelize, and
 	optimize the computation of visualizations and obtain interactive speeds
 	despite large datasets. Vizketches are executed in a tree, where leafs
 	process shards in parallel and merge results toward the root.}
 \label{fig:intro}
 \label{fig:arch}
\end{figure}

To render visualizations quickly, \sysname introduces a new distributed
  engine to compute vizketches (Fig.\ref{fig:intro}).
Clients access the system via a user interface in a web browser (top of figure),
  while the dataset is partitioned across a set of worker servers (bottom).
The user interface triggers a visualization, such as a histogram on a chosen column.
To produce the visualization, the system executes two phases: preparation and rendering.
The \emph{preparation phase} computes broad parameters required to
  produce a proper visualization---for example, a histogram needs
  to find the data range and number of items to determine
  appropriate bucket boundaries and sampling rates.
Next, the \emph{rendering phase} computes the values required for the
  visualization---for example, the height of each histogram bar.
This phase utilizes a vizketch to compute
  with the minimum accuracy for a good visualization.  
The rendering phase produces partial results that incrementally
  update the visualizations, so the client sees an initial visualization
  quickly and subsequently sees more precise results.
Both preparation and rendering phases use an execution tree
   to distribute the computation across the workers.
The engine provides other important functionality that we
  describe in the paper: caching
  computations, distributed garbage-collection,
  and failure recovery.
Furthermore, the engine has a modular design that
  allows developers to add visualizations easily using new
  vizketches without dealing with concurrency, communication, and
  without needing to understand the structure of an existing query
  optimization engine; in practice support for a new storage layer or
  for a new visualization type can be added in a couple of person-days of work.
  
The engine of \sysname  differs fundamentally from general-purpose query engines in 
  two important ways. 
First, due to the characteristics of vizketches, \sysname queries are scalable by construction: more specifically, queries are guaranteed to run in time $O(n/c)$, produce results of length $O(\log n)$, using memory of size $O(\log n)$ where $n$ is the number of elements in the dataset and $c$ is the number of worker cores\footnote{Assuming a balanced partition of the data between workers.}. In addition, many queries run in time $O(1)$.
Second, \sysname produces compact results designed to be rendered efficiently on the screen. 
By contrast, general-purpose engines are not concerned about efficiency renderings; 
their queries could produce large results that take longer to visualize than to compute (\eg, 
returning billion points to be plotted)~\cite{christopher-tableau19,Abraham-vldb13,Splunk}.

We evaluate \sysname and its vizketches.
We find that \sysname can support tables with 1.4 trillion cells
  while providing fast response.
With this scale and data in memory, operations take 1--15 seconds.
\sysname displays an initial partial views even faster, which
  is incrementally updated until it converges to the final view.
With cold data read from an SSD, operations take 2--24 seconds, and an
   initial view still appears within seconds.
For datasets with hundreds of billions of cells, \sysname computes
  complete answers in under a second for most queries.
This is faster than the current approach of connecting
  a visualization front-end to a general-purpose analytics back-end.
We also find that \sysname has broad functionality for answering a wide range of
  questions.
Vizketches are an order of magnitude faster than a
  popular commercial in-memory database system to compute histograms; and their
  performance scales linearly or
  sometimes super-linearly with the number of threads and servers.

To demonstrate the usability of \sysname, we provide a short video and
a live demo running on AWS using small EC2 instances (these links are
also available in our github repository):
  \vspace{1mm}

{\small\noindent
Video:\url{https://1drv.ms/v/s!AlywK8G1COQ_jeRQatBqla3tvgk4FQ} \\
Demo:\url{http://ec2-18-217-136-170.us-east-2.compute.amazonaws.com:8080}
}

\vspace{1mm}
In summary, in this paper we propose \sysname, a spreadsheet for big data.
\sysname makes two novel contributions.
First, it introduces vizketches, an idea that combines mergeable summaries
with visualization principles; we
give vizketches for each chart and tabular view in \sysname, by finding
appropriate mergeable summaries and parameters to render
  information efficiently yet provably accurately.
Second, \sysname demonstrates how to efficiently compute vizketches by introducing
  a new scalable distributed analytics engine that
  caches computations, performs distributed garbage-collection, and handles failure recovery, while
  achieving the scalability and speed required for an interactive spreadsheet.

While the above contributions are pragmatic, we believe this work also contains
  a fundamental contribution.
We raise and defend two hypotheses: (1) mergeable summaries are powerful enough to
  efficiently and accurately visualize massive datasets, and (2)
  spreadsheets can significantly benefit from a specialized engine.
\sysname demonstrates these hypotheses
  empirically by giving vizketches for many visualizations, by building
  an engine for vizketches, and by quantifying its benefits.
\sysname is an open-source system with an Apache 2 license, available
at \url{https://github.com/vmware/hillview}.

Due to space limitations, we provide an extended version of this paper~\cite{extended}, with additional details:
a formal computational model that captures vizketches, formal definitions of correctness and efficiency,
detailed descriptions of vizketches, and correctness proofs.

\pagebreak

\section{Why a new engine}\label{sec:engine}

In a famous paper, Stonebraker et al. advocate
  for designing database systems targeted for specific domains, because doing so
  can dramatically improve performance over one-size-fits-all solutions~\cite{endofarchitectural}.
This approach has worked well for several domains:
   data warehousing, stream processing,
   text, scientific, online transaction processing, etc.
More recently, Fisher~\cite{Fisher-hilda16} and Wu et al~\cite{Wu-vldb14}
  point to the need for collaboration
  between visualization and data management systems.
\sysname arises from these insights: we apply the database specialization approach
  to big data spreadsheets, where existing
  solutions fall short in scale and performance.

\sysname raises an important question.
Data analysts may want to apply rich pipelines to data involving different frameworks, tools,
  and programming languages.
For example, they may use a statistical package in R,
  then apply a machine learning algorithm in C++, followed by
  some hand-written scripts in python.
How can \sysname integrate in this environment, given \sysname's specialized engine?
  

\sysname addresses this concern by
  adopting a versatile data layer that can connect to
  other tools in the pipeline.
In particular, \sysname can operate directly on data stored in SQL databases,
  NoSQL systems, JSON files, CSV files, columnar-oriented
  files such as Parquet or ORC, and other big-data systems (Hadoop+Spark, Impala),
  without any data transformation overheads.
This is because \sysname does not require data ingestion to produce indexes,
  or repartition data: the efficiency of
  vizketches permits \sysname to operate on raw data partitioned horizontally in
  arbitrary ways across servers: there are no requirements that partitions
  contain contiguous intervals or specific hash values.
The only requirements of the data layout is that
  (1) data be horizontally partitioned ideally with approximately equal-sized
  partitions available to each worker, and
  (2) data does not change while \sysname is
running\footnote{This requirement is common in data warehousing and analytics systems.}.
The latter requirement can be met by using a data layer that provides snapshots, immutable data,
   or by pausing data modifications while \sysname runs.
If a processing pipeline meets these requirements, then it is easy
  to insert \sysname into the pipeline.
For example, we can connect the output of a batch-processing
  system to \sysname for exploration, and then output
  \sysname visualizations as data files or images that are processed by
  subsequent tools in the pipeline.

  

\section{Goals and requirements}\label{sec:goals}

Our main goal is to develop a big data spreadsheet.
As a data analytics tool, we are interested in functionality
  to explore and summarize data, such as navigation, selection,
  and charts.
These are mostly read-only operations---our tool is for analytics exploration
  rather than transaction processing, data wrangling, cleaning, etc.
So, we are less interested in providing interactive editing functionality,
  but we wish to provide ways to compute new columns from existing ones
  (\eg, compute a ratio of two columns).
We now explain our requirements in more detail.

\subsection{Why trillions of cells}

Even small and medium companies can generate a trillions cells of data.
These companies collect data over time from their
  servers, where each server might produce logs and metrics hundreds of times per
   minute, and a data center could have dozens of such servers.
For example, 50 servers logging 100 columns at a rate of
100 rows per minute generate in a month 21.6B cells on 216M rows,
or 1T cells and 10B rows in 46 months.

\vfill
\subsection{Environment}

We target an enterprise computing environment, with
  tens of commodity server machines in a rack hosted
  in a private or public cloud.
We want to use as few servers
  as possible, as most companies cannot afford
  thousands of servers to run a spreadsheet.

\subsection{Tabular views functionality}\label{sec:tabular}

At first thought, it is unclear what a spreadsheet with a billion rows
  should do.
Clearly, paging through all rows is ineffective, but analysts
  may wish to find patterns and then inspect individual rows.

In our experience browsing big data, we found that
a spreadsheet must support at least the following functionality.

\begin{itemize}
\item Select data based on rich
  criteria to produce fewer rows
  (\eg, rows with timestamps in the past hour).

\item Select columns to show
  (\eg, date and server).

\item Sort by a set of columns
  (\eg, date first, server next).

\item Aggregate duplicates and show repetition counts
  (\eg,  selecting just date and server could create millions of
  repetitions: all entries produced by each server on each day).

\item Search free-form text (\eg, server Gandalf) by exact match,
    substring, regular expressions, case sensitivity, etc.

\item Move a page forward or backward.

\item Scroll forward and backward using a scroll bar.

\item Extract features using tools such as
  heavy hitters (finds most frequent elements) and
  Principal Component Analysis~\cite{pca-14}.
\end{itemize}

We consider whether this functionality suffices in
  \S\ref{sec:casestudy}, but we
expect the list will grow over time, much like conventional spreadsheets have
  evolved, so we also need a flexible framework that allows us to extend
  the system.

\subsection{Visualization functionality}\label{sec:visualizations}

\begin{figure*}
{
\begin{tabular}{lp{2.5in}l}
 {\bf Name}          & {\bf What it shows}           & {\bf Example} \\
 CDF                 & Distribution of variable  & \# events before noon \\
 Histogram           & Frequency of variable for each bucket & \# events per hour of day \\

 Stacked histogram   & Frequency of first variable
                       and frequency of second variable
                       grouped by first              & \# events of each type per hour of day \\
 Normalized stacked hist.& Ditto but bars normalized         & \% of events of each type per
                                                           hour of day \\
 Heat map            & Frequency of two variables        & \# events for each server and hour \\
 Trellis plots       & Arrays of the other plots grouped
                       by one or two variables           & \# events for each server and hour,
                                                           for each datacenter \\
\end{tabular}

\begin{center}
\begin{tabular}{@{}cc@{}}
	\begin{minipage}{3.25in}
          \includegraphics[width=3.25in]{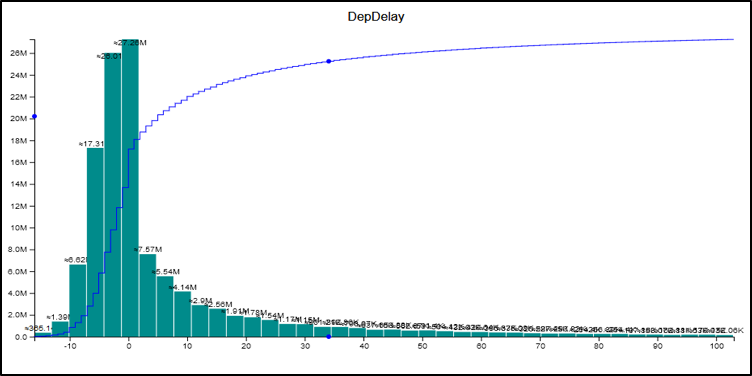}
	  \centerline{Histogram and CDF}
    \end{minipage} &
	\begin{minipage}{3.25in}
          \includegraphics[width=3.25in]{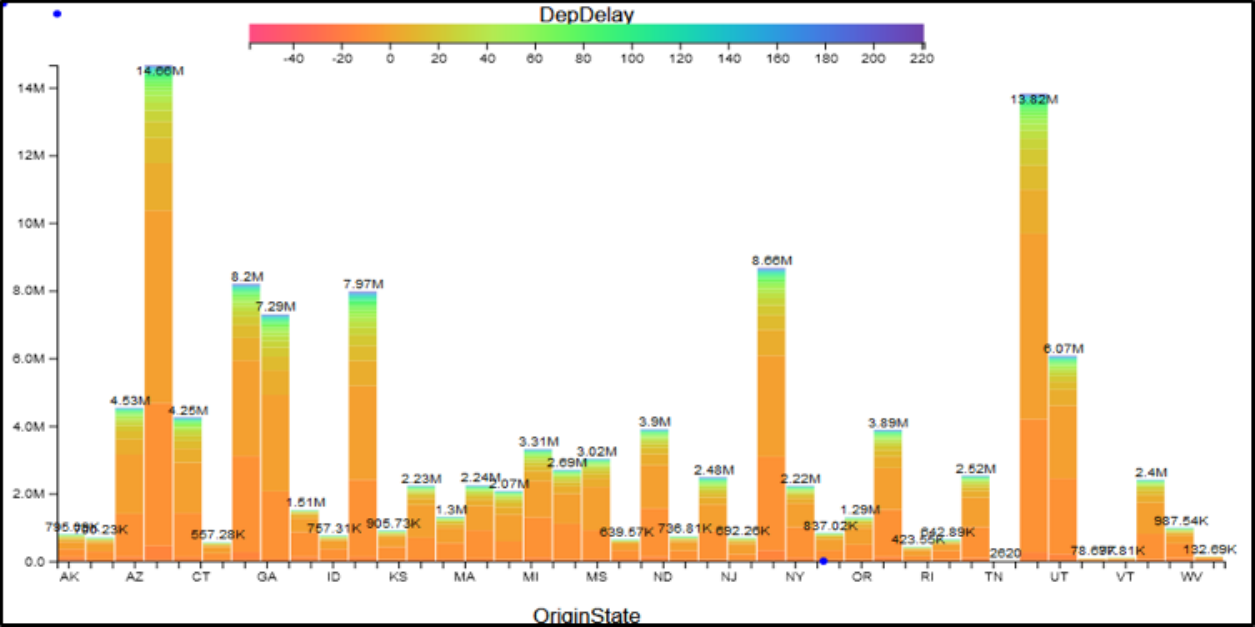}
	  \centerline{Stacked histogram}
    \end{minipage} \\ [4mm]

	\begin{minipage}{3.25in}
          \includegraphics[width=3in]{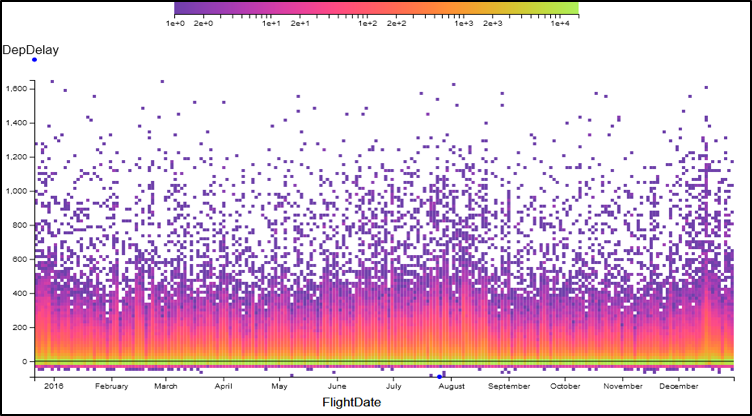}
	  \centerline{Heat map}
    \end{minipage} &
    \begin{minipage}{3.25in}
          \includegraphics[width=3.25in]{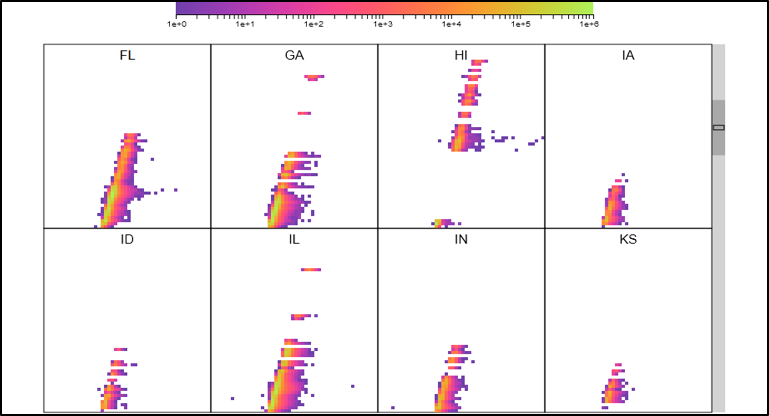}
	  \centerline{Trellis plot with heat maps}
    \end{minipage} \\
\end{tabular}
\end{center}
}
\caption{Some clutter-free visualizations for large datasets.
	Visualizations cover a single variable (column) or
	multiple variables, up to four.
}
\label{fig:visualizations}
\end{figure*}

We are also interested in obtaining various visualizations of
  columns we choose.
But we face a problem with big data: graphs with billions
  of points can produce useless black blobs and other clutter.
We want to support
  visualizations that can avoid this problem~\cite{shneiderman-sigmod08,elmqvist-tvcg10}, such
  as histograms, stacked histograms, and heat maps
  (Figure~\ref{fig:visualizations}).
These visualizations generalize charts, such as
  x-y plots and bar charts (subsumed by heat maps);
  and pie charts (subsumed by heavy hitters (\S\ref{sec:tabular})).
We also want to extend the system with future new visualizations.

For each visualization, we want to
  inspect the value of individual points, change parameters (\eg, \# buckets in histogram) and,
  if applicable, understand trends, correlations, and swap axes.
Furthermore, we want to zoom in parts of the data, by
  regenerating the visualization for a subset of its data as determined by a mouse selection.


\subsection{Other features}

\xparagraph{Data types.} We want to support integers, floating-point numbers,
  dates, free-form text, and
  strings describing categorical data.

\xparagraph{Map functions.}
We want to produce a new column by combining existing ones using
user-defined map functions (\eg, a ratio of two columns).

\xparagraph{Data sources.}
We want to read data from a variety of common sources (comma-separated files, SQL databases, row columnar files such as ORC, future formats, etc).

%
%

\section{Vizketches}\label{sec:viz}

Key to providing the required performance of \sysname,
  vizketches are a simple concept that
  combine the idea of mergeable summaries (or sketches) from the algorithms
  community with the principle of visualization-driven computation from the graphics
  community.

\subsection{Background} \label{sec:background}

\xparagraph{Mergeable summaries.}
Intuitively, a summarization method computes a compact representation (``summary'')
  of a large dataset, which can then answer approximate queries
  on the dataset.
A summarization method is mergeable~\cite{Agarwal2012} if the summary can be
  obtained by merging many summaries computed independently
  over parts of the dataset.
More precisely, a mergeable summarization method consists of two functions
  $\t{summarize}(D)$ and $\t{merge}(S,S')$.
The first takes a dataset $D$ and returns a summary; the second
  merges two summaries and returns another summary.
A summary is small compared to $D$---typically by many orders of
  magnitude---and it can approximate queries on $D$ (the allowable
  queries depend on the choice of summarization method).
Summaries
  of two separate datasets can be merged via the $\t{merge}$ function:
  \[ \t{summarize}(D_1 \uplus D_2) =  \t{merge}(\t{summarize}(D_1),
  \t{summarize}(D_2)) \]
  where $D_1$ and $D_2$ are mutisets and
  $\uplus$ is multiset union.
There are summarization methods for many types of queries, such as
  histograms, heavy hitters, heat maps, and PCA.
Many summarization methods are sketches from the streaming
  algorithms literature~\cite{cormode-cacm17},
  and so the community sometimes mixes these two concepts.
However, a summarization method can also use sampling, which
  can be more efficient because it does not scan all data.
The summarization method has two accuracy parameters:
  an error $\epsilon$ and
  an error probability $\delta$, with the guarantee that
  an approximation computed from a summary has
  error at most $\epsilon$ with probability $1-\delta$.
For a more formal description of our computational model, we refer the reader to Appendix~A
  of the extended version of this paper~\cite{extended}.

\xparagraph{Visualization-driven computation.}
In computer graphics, rendering is an expensive operation
  that must be optimized.
To do that, a basic principle is to drive the computation based on
  what will be visualized and its resolution, taking into
  consideration the limits of human perception and
  the lossy channels of displays.
This principle is behind many graphics techniques, such as
  ray tracing, culling, and imposters~\cite{computergraphicstextbook}.

\subsection{Basic idea}\label{sec:basicidea}

A vizketch is a mergeable summary designed to
  produce a good visualization.
More precisely, a vizketch method targets a specific visualization
  (\eg, a histogram)
  with a given display dimension (width and height in pixels).
The vizketch method consists of the two functions of
  a mergeable summary, \t{summarize} and \t{merge}, with
  parameters carefully chosen to achieve two goals:
  the summary is \emph{small}, and
  it permits a \emph{good rendering} of the visualization.

{\em Small summary} means that its size depends only on the length of the description
    of the visualization, not on the input size.
More precisely, visualizations are inherently limited by the finiteness of their renderings, so they
  have a short description (\eg,
  a histogram is described by its bucket boundaries and heights).
The length of this description is a lower bound on the size of the summary.
We seek summaries whose size is polynomial in this length, rather than
  the data set size.
The key hypothesis behind \sysname is that visualizations
   always admit vizketches with such small summaries.
This hypothesis is not obvious; it can be formalized with proper
  definitions of the computational
  model, visualizations, etc., but this is outside the scope
  of this paper.
Instead, \sysname supports this hypothesis empirically: we give vizketches for many visualizations, by adapting techniques from the
  sketching/streaming literature.

{\em Good rendering} means two things.
First, the rendering has a bounded error with high probability (\eg, histogram
  bars are off by at most 1 pixel).
Second, the rendering is not cluttered (\eg, there are at most 50 buckets
  for a histogram when the screen width is 200 pixels).
The precise requirements are carefully chosen for each type of visualization.
These choices are made so that a human can consume the information effectively
  without perceiving any errors in the approximation.


To use vizketches, \sysname defines a computation tree whose nodes are assigned
   to the servers (Figure~\ref{fig:arch}).
   \sysname assumes that the data is stored on a distributed storage layer, and is
   sharded into small chunks, which are distributed to the
   tree leaves.
The sharding can be arbitrary: chunks need not be sorted or partitioned by a specific key.

To perform a visualization, each leaf independently runs the vizketch's
  \t{summarize} function on the shards that it has; this function might
  choose to sample or scan the data in the chunk\footnote{This choice can be made independently for each chunk.}.
The summaries are then merged along the computation tree, using the
  vizketch's \t{merge} function.
The root receives the final summary, which reflects a view of the entire dataset
  and produces the rendering of the visualization for the client.

Vizketches parallelize the computation across threads and servers,
  while reducing computation and network bandwidth to only what is necessary for
  a good rendering.
They can also provide partial results for progressive visualizations, in addition
  to other benefits (\S\ref{sec:benefits}).
We now describe specific vizketches.

\subsection{Algorithms}\label{sec:vizchart}

\sysname uses a large number of vizketches.
Some produce graphs (histograms, stacked histograms, heat maps, trellis plot);
  others produce information for the spreadsheet tabular view (next items,
  quantile for scroll bar, find text, heavy hitters).  We describe a few here;
others are omitted due to
  space limitations but they follow a similar approach and
  can be found in Appendix~B of~\cite{extended}.
Vizketches have rigorous guarantees of correctness, which
  we present in Appendix~C of~\cite{extended}.
  
A vizketch is parameterized by the target display resolution, and
  produces calculations that are just precise enough to render at that resolution.

\begin{figure}
\begin{tabular}{@{}c@{\hskip 10mm}c@{}}
\includegraphics{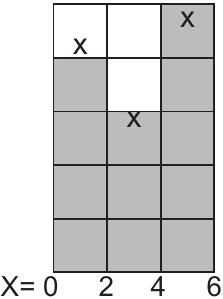} &
\includegraphics{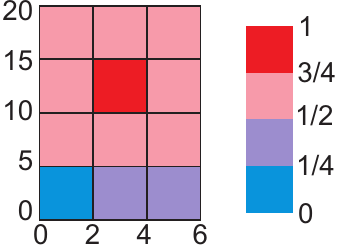}
\\
(a) & (b) \\
\end{tabular}
\caption{Charts in \sysname have an error of at most 1/2 pixel or one color shade
	     with high probability.
	(a) A histogram with three bars. The ${\times}$ indicates the correct height for the bar
	    at most one 1/2 pixel away from the rendering.
        (b) A heat map (left) and the density color map (right).
            The x-axis has bins for the first variable; the y-axis, for the second variable.
	    The color indicates the density of each bin, where the error is at most one color shade
	    with high probability.
}
\label{fig:viz}
\end{figure}

\xparagraph{Histograms.}
We are given a numerical column (or a value that can be readily converted to a real number, such as a date)
with range $[x_0,x_1)$, a number $B$ of histogram bars, and
their maximum pixel height $V$.
The histogram vizketch (Figure~\ref{fig:viz}(b)) divides the range $[x_0, x_1)$ into $B$ equi-sized intervals, one per bin.
To maximize use of screen, we should scale the bars so that the largest one
  has $V$ pixels.  We furthermore allow an error of .5 pixels in the estimation of the height of each bar.
We prove in Appendix~C of~\cite{extended} that to obtain this error with probability $< 1-\delta$,
it is sufficient to sample $n=O(V^2 B^2 \log(1/\delta))$ items from the dataset.  Notice that this function
is independent on the dataset size, and depends only on the screen size.
The \t{summarize} function outputs a vector of $B$ bin counts, and the
  \t{merge} function adds two vectors.

\def\ncolors{20}

\xparagraph{Heat map.}
We are given two columns $X$ and $Y$
  with ranges $[x_0,x_1)$ and $[y_0,y_1)$, and the pixel dimensions $H \times V$.
A heat map (Figure~\ref{fig:viz}(d)) defines bins in two dimensions, where each bin consumes $b \times b$ pixels ($b$ is 2 or 3, depending on the screen resolution).
Thus, we have $B_x = H/b$ and $B_y = V/b$ bins for $X$ and $Y$.
The density of the data in a bin is represented by a color scale.  If we use $c{\approx}20$ distinct colors,
the required accuracy for each bin density is $1/2c$.
This requires a target sample size $n=O(c^2 B_x^2 B_y^2 \log(1/\delta))$\footnote{Sampling can be used only
if the map from count to color is linear.}.
The \t{summarize} function samples data with the target rate, counting the number of values
  that fall in each bin.
It outputs a matrix of $B_x \times B_y$ bin counts.
The \t{merge} function adds two such matrices.

\xparagraph{Next items.}
This vizketch is used to render a tabular view of the spreadsheet given the current row shown at the top $R$
  (or $R=\bot$ to choose the beginning of the dataset).
We are also given a column sort order, and the number $K$ of rows to show.
This vizketch returns the contents of the $K$ distinct rows that follow $R$ in the sort order.
The \t{summarize} function scans the dataset and keeps a priority heap with the $K$
next values following row $R$ in the sort order.
The \t{merge} function combines the two priority heaps by selecting the
smallest $K$ elements and dropping the rest.

\xparagraph{Heavy hitters.}
A vizketch to find heavy hitters works by sampling.
 Let $K$ be the maximum number of heavy hitters desired.
The basic idea is to sample with a target size $n$ (determined below),
  and select an item as a heavy hitter if it occurs with frequency at least $3n/4K$.
A statistical calculation shows that by picking $n=K^2 \log(K/\delta)$,
  with probability $1-\delta$
  we can obtain all elements that occur more than $1/K$ of the time and no
  elements that occur fewer than $1/4K$ of the time.
  This method is particularly efficient if $K$ is small.  We employ several other
  algorithsm for finding heavy hitters, described in Appendix C of~\cite{extended}.

\vfill
\subsection{Benefits}\label{sec:benefits}

Vizketches bring many benefits to \sysname.
In the list below, the parentheses indicate from where the benefit is inherited:
  S means sketches/mergeable summaries, V
  means visualization-driven computation, and S+V means the combination of both.

\begin{itemize}
	\item {\em Parallel computation (S).} Servers and cores within servers
	independently compute on different parts of the data, and the result is merged.

	\item {\em Bandwidth efficiency (S+V).}  When a server finishes its computation,
	it communicates only a compact summary to be merged.

	\item {\em Computation efficiency (S+V).} Some computations can done over
	a small sample of data based on the required accuracy.

	\item {\em Progressive visualization (S).} As servers complete their computation,
	the system computes a partial summary that gradually progresses to the
	final result. This ensures that slow servers and tail latencies
	do not hinder interactivity. Users can cancel a visualization after seeing partial results.

	\item {\em Accurate visualization (S+V).} The resulting visualization has a precise
	  accuracy guarantee.

	\item {\em Scalability (S+V).} As we add more data, vizketches can sample more
	aggressively to enhance efficiency while achieving the required
	accuracy.

	\item {\em Easy to obtain (S).} There is a rich literature on mergeable
	summarization methods and sketches of various types (histograms, heat maps,
	heavy hitters, etc); these sketches can often be converted into vizketches through
	a relatively simple analysis that translates the accuracy of the sketch into the
	required accuracy of the visualization, as illustrated above.

	\item {\em Modularity (S).} New visualizations can be added to \sysname by defining new
	vizketches as two simple functions (\S\ref{sec:background}) without the developer
	worrying about distributed systems aspects.
\end{itemize}

\section{Design and architecture}\label{sec:design}

We now explain in detail the design and architecture of \sysname,
  starting with its high-level design choices (\S\ref{sec:choices}), followed
  by a detailed description in the subsequent sections.

\subsection{Design choices} \label{sec:choices}

We now explain the key design choices of \sysname, which derive from the power
  and characteristics of vizketches.

\begin{itemize}

\item {\em Distribute computation while minimizing server coordination.}
To answer a query, \sysname launches a computation tree to 
  efficiently distribute the query to worker servers and aggregate the results
  according to the vizketch computations.

\item {\em Storage-independence.} \sysname can access data in a wide variety of formats (SQL, NoSQL, text, JSON, etc),
  with few restrictions on how data is partitioned (\S\ref{sec:engine}), and without the
  need to pre-compute indexes or perform extract-transform-load.
As a result, \sysname does not require any pre-processing to ingest data.
This is beneficial to integrate \sysname into a diverse analytics pipeline (as explained in \S\ref{sec:engine}),
  and this is possible because the efficiency and parallelization of vizketches permits
  forgoing data conversions, repartitioning, and pre-computations.
  
\item {\em Sample data in a controlled manner.} Sampling improves efficiency but introduces error.
Vizketches allow \sysname to sample while bounding the error to what we can perceive.

\item {\em Modular algorithms.} Programmers who write
  vizketch algorithms do not have to worry about concurrency, communication, or
  fault-tolerance; they just implement single-threaded code, and the
  architecture handles all such issues in a uniform and transparent manner.

\end{itemize}

\subsection{Architecture} \label{sec:arch}

Figure~\ref{fig:arch} shows the architecture of \sysname.
\sysname is designed as a cloud service accessible
  to clients through a web interface.
A web browser handles user interaction with the spreadsheet
  and renders the the charts incrementally as computation results arrive.
To produce a visualization, a web server launches the
  required computation as one or more {\em execution trees}.
Communication happens only along the edges of
  the tree, and is restricted to small messages:
  queries in one direction and summaries in the other.
Each tree is
 rooted at the web server,
  followed by one or more layers of aggregation nodes,
  and several leaf nodes.
The leaf nodes perform the actual computation over disjoint
  partitions of the dataset.
These nodes have an in-memory data cache in front of the
  data in repositories.
There is also a computation cache to
  reuse prior computations.   
The aggregation nodes are intended to scale the system to
  many servers; a small deployment with
  tens of servers needs only one layer.  

\subsection{Execution tree}\label{sec:exectree}

A visualization typically involves two execution trees, each
  intrinsically linked to a mergeable summary.
The first tree computes data-wide parameters such as
  the size and range of the data set; this computation may
  be cached from previous visualizations.
The second tree computes a vizketch for the visualization
  with the required accuracy based on the results produced by the
  first execution tree.
  
The execution of each tree is based on the
  \t{summarize} and \t{merge} functions (\S\ref{sec:basicidea})
  of the mergeable summary.
A tree executes in two phases.

The first phase initiates the computation from the root
  down the tree to each leaf, and causes the leaf nodes
  to apply the \t{summarize} method on their data partition.
To parallelize execution within a server, each server
  runs multiple leaf nodes: there is a thread pool that
  serves leafs with work to do.
To facilitate this process, the data partition within a server
  is divided into micropartitions of 10-20M rows, each
  micropartition assigned to a leaf.
  
The second phase, in its most basic form, executes from the leafs toward the
  root, causing each node to aggregate results from its
  children through the \t{merge} method.
Thus, ultimately the root node combines the output
  of all nodes, and the result can be rendered.
When processing large datasets in a distributed system, there
  may be variation in the processing times across servers and partitions.
If the root had to wait
  for all other nodes to finish, its completion would be disrupted
  by any stragglers, affecting the interactive experience of users.
To address this problem, nodes periodically propagate partially
  merged results
  of the vizketch without waiting for all children to respond.
Thus, the root receives partial results and sends them to the client UI,
  before it gets the final results.
The web browser then renders results as they arrive, so that
  users can see a progression of the computation.
\sysname shows a progress bar that reflects the number of leafs
  that have completed.
Users can cancel the computation based on the partial
  results they see.
  
There is a trade-off between the freshness of the partial results and
the bandwidth savings produced by aggregating partial results.  After
receiving a result from a child node, aggregation nodes wait for 0.1
seconds and aggregate all results that arrive within this interval.
This provides frequent updates to the UI; the increase in
communication costs is modest because all vizketch results are small
by construction.
  
\sysname allows users to cancel computations (\eg, because a partial
  visualization is satisfactory).
This is done by interrupting an execution tree with a high priority cancellation message that
  bypasses the queuing mechanisms in the communication between tree nodes.
This message causes tree nodes to do two things:
  remove work for that computation that was previously enqueued, and
  ignore requests for micropartitions not yet started.
We currently do not stop ongoing computations on a micropartition.

\subsection{Data input, caching, and data output}

Unlike most database systems, \sysname reads data repositories without pre-processing,
  repartitioning, or other optimizations.
This is possible because the computational engine of
  \sysname---based on vizketches---makes few assumptions about the data.
The assumptions are that repositories do not change while they
  are accessed (this can be provided by using storage snapshots or controlling write access)
  and data is horizontally partitioned, ideally with approximately equal-sized
  partitions available to each worker, so that data can be read in parallel.
When a worker needs a column, it reads it completely from the data repository taking
  advantage of fast sequential access and
  columnar access if the repository supports it (SQL, Parquet, ORC).
  Once data is read, it is kept in an in-memory cache; the cache purges entries
  not used for a while (currently 2 hours).

\sysname uses two types of caching: data and computation.
The first is an in-memory cache of the raw data in the data repositories.
The data cache is organized by column to provide data locality, since
  vizketches tend to operate on relatively few columns.

The computation cache stores results produced by mergeable summaries;
  these results are small, allowing a large number of results to be cached.
This is useful for mergeable summaries that provide
  auxiliary functionality, such as column
  statistics, which are used repeatedly and are deterministic.
The computation cache is indexed by what mergeable summary was used
  and what dataset was operated on.
  
\sysname can save a derived table (\S\ref{sec:deriving}) to a data repository,
  by having each worker store its partition of the data.
This is implemented through a special vizketch with a summarize function that
  writes a data record to the repository and returns an error indication, while
  the merge function combines error indications.

\subsection{Vizketch modularity and extensibility}

The inherent structure of vizketches permits
  \sysname to cleanly separate them from the rest of its architecture
  so that developers can implement new vizketches without
  the hard concerns of distributed systems (communication,
  coordination, fault tolerance, etc) or data storage.
Specifically, to support a new vizketch, a developer needs to implement the following things:
  (1) a serializable\footnote{I.e., the type should have a serialization method to convert an instance into a byte sequence for network transmission.} type for the vizketch summary,
  (2) an implementation of the \t{summarize} and \t{merge} functions of the vizketch; these all operate on the in-memory columnar representation of the data, and are independent on the storage layer,
  (3) code to render the vizketch summary as a visualization in the user interface of the spreadsheet in the browser,
  (4) code to trigger the visualization through a user interface action, and
  (5) a function to connect the user interface action to the invocation of the vizketch in the root node.
None of these functions are concerned with concurrency (they are single-threaded), and
  most of them can be implemented with only tens of lines of code---the sole exception is (3), which
  requires more code to provide the graphical functionality.  We quantify the effort to for step (2) in \S\ref{sec:eval:ease}.

  \vfill
\subsection{Data transformations}\label{sec:filter}\label{sec:deriving}

Users may wish to generate new data from existing data
  as part of the data exploration process.
Users can do that externally to \sysname through other analytics
  tools, and then import the results into \sysname for inspection
  (\S\ref{sec:engine}).
Alternatively, \sysname provides some support for deriving new
  data through two common operations: selection (filtering) and user-defined map operations (\S\ref{sec:goals}).

Selection permits a user to create a new table that contains a subset of the
  rows of another table (\eg, rows where the year column is 2019).
A particularly useful selection operation in a spreadsheet is to zoom in part of a graph,
  which corresponds to choosing the rows within the zoom window. 
To provide this functionality, \sysname allows mergeable summaries
  to work on subsets of rows of the dataset.
More precisely, a table can be derived from other tables by choosing
  a subset of the rows.
To save space, the tables share common data and
  store a ``membership set'' data structure that identifies which
  rows are contained in the table.
This membership set data structure has different implementations, depending
  on the density of the filtered data.
Dense tables that contain most rows store a bitmap, while
  sparse tables store a hashset of the rows indexes.
  This information is kept locally for each data partition.

When executing the summarize method, some vizketches
  work by sampling a subset of rows.
We must ensure that sampling is efficient (it does not
  require reading each row) but it is also correct (it
  picks rows uniformly at random).
For sparse tables, we generate the first sample by
  choosing a random row number for the first element; we generate
  the following samples by returning the next elements
  in sorted order of their hash values.
For dense tables we walk randomly the bitmap in increasing index order.

User-defined maps permits a user to create a column from existing
  ones (\eg, add two columns), where the map 
  is an arbitrary function.  Some map functions are built-in (e.g., converting
  strings to integers); additional functions can be written by
  users in Javascript.
To support this functionality, \sysname creates a new table
  with the new column populated by running the map function
  at the leafs of the execution tree.
Currently, this data is stored only in the in-memory caches;
  if the cached data is reclaimed, the column is recomputed on demand.
We believe this is a reasonable choice for a spreadsheet, since
  derived columns tend to be short-lived.

\subsection{Memory management}\label{sec:memory}

Early versions of \sysname used a distributed garbage-collection
protocol to handle memory management.  This protocol was complex and
fragile (for example, loss of network messages could cause memory
leaks).  In the current version we have simplified memory management
by aggressively using only soft state: all in-memory data structures
are disposable, including at leaf-, aggregation- and root nodes.  The
only requirement to implement this architecture is for the storage
layer to provide an API to read a particular snapshot of a dataset; in
this way, in-memory data is reconstructed by reloading the original
snapshot.  We use the Partitioned Data Set architecture from
Sketch~\cite{budiu-egpgv16} to represent distributed objects; unlike
sketch, all remote references are ``soft'' --- they may not point to
valid data structures.

Each machine performs independently garbage-collection; a caching
layer maintains a working set of recently accessed objects in memory.
In-memory cached objects at leaf nodes can be reconstructed by reading
data from disk; tables obtained from filtering (\S\ref{sec:filter}) or
by deriving new columns (\S\ref{sec:deriving}) can be regenerated by
re-executing the operation that created them in the first place.

When the root node attempts to access a remote object on a leaf which
no longer exists the leaf reports an error.  The root node then
re-executes the query that produced the missing object.  This may
require re-executing other queries, that produced the source objects;
the recursion ends when data is read from disk.

To enable query re-execution, the root node maintains a redo log with
all executed operations.  The redo log is
the only persistent data structure maintained by \sysname (recall that
the storage layer is not part of \sysname).

\subsection{Fault tolerance}

\sysname provides fault tolerance by logging operations that
  initiate each execution tree, and
  lazily replaying operations to reconstruct node state.
When the root node restarts after a failure, it reads the redo log to memory, but does not replay it yet.
Replaying occurs only when the user tries to access a dataset that no longer exists, as described in \S\ref{sec:memory}.

Worker nodes are stateless, so restarting the node after a failure
  is equivalent to deleting all cached datasets.  These datasets are
reconstructed by the root node on demand by replaying log operations.
  
This lazy aproach is suitable for a spreadsheet, because most views are
  short-lived results that a user never accesses again.


For this replay mechanism to work, vizketches must be deterministic,
  otherwise a restarted node becomes inconsistent with nodes that never
  crashed.
To provide determinism for randomized vizketches (\eg, those that
  use sampling), the log includes the seed used for randomization.

\section{Implementation}

\sysname consists of 35000 lines of Java and 16000 lines of TypeScript code.
The user interface in the browser is implemented in
TypeScript~\cite{typescript}, using parts of the D3 JavaScript
library~\cite{d3-2011}.  Graphics is done using SVG~\cite{svg-w3c11}.
The web server runs the Apache Tomcat application server~\cite{tomcat}.
The browser gets progressive replies from web server
  using a streaming RPC based on Web Sockets~\cite{websockets11}; these
RPC messages are serialized as JSON.
%
The cloud service is implemented in Java.  We use Java's type-safe
object serialization facilities for sending queries and data between
machines. We use the fast collections Java library~\cite{fastutil} for
efficient data structures, with customizations for faster sampling.
For server-side JavaScript we use Oracle Nashorn~\cite{Nashorn}.

We use a variety of open-source libraries to interface with external
storage layers (e.g., csv files, various log formats (e.g., RFC
5424), JDBC connectors, columnar binary formats such as Parquet or
ORC, etc).
%
%
The communication between back-end machines uses GRPC~\cite{grpc}.
The core communication APIs are based on reactive streams, using
RxJava~\cite{reactivex, Meijer-queue2012}.  We use RxJava's
\texttt{Observable} datatype for many purposes:
(1) It represents a stream of partial results, (2) it offers
support for operation cancellation, through its \texttt{unsubscribe}
method, (3) it is used for reporting progress to the user for
long-running operations, displayed in the form of progress bars, and
(4) it is responsible for managing concurrent execution on multi-core
machines (using the \texttt{observeOn(threadPool)}; this thread pool
is used for all of the workers' computations.
The in-memory tables use as much as possible Java arrays of base types
to reduce pressure on the Java GC.  String columns use dictionary
encoding for compression.

\section{Evaluation}

Our evaluation goal is to determine whether \sysname provides interactive
  performance with large data sets, how \sysname compares to existing systems,
  how vizketches contribute to that goal, and
  how effective the spreadsheet is.

\xparagraph{Summary.}
We find the following results:

\begin{itemize}
\item \sysname can handle spreadsheets with 130B rows and 1.4T cells
  using only 8 servers.
   At the upper range, visualizations can take 20s when loading from disk,
   but the first partial visualization appears in a few seconds and gets
   gradually updated. This is much better than existing systems
   (\S\ref{sec:eval:macro}).  For smaller datasets most response times are
   on the order of hundreds of milliseconds.
 \item Vizketches perform well on a single thread and scale well with the
   number of threads and servers. Vizketches based on sampling scale
   super-linearly. This performs signficantly better than a database
   system (\S\ref{sec:eval:micro}).
 \item Vizketches are key in \sysname: they implement a broad range of functionality 
   of the spreadsheet, 
   to the extent that they are the sole way to access data in the system (\S\ref{sec:eval:applic}).
 \item Vizketches are easy to code and do not require an understanding of distributed
   systems (\S\ref{sec:eval:ease}).
 \item \sysname is a spreadsheet with many useful features, able to answer
   a diverse set of queries effectively (\S\ref{sec:casestudy}).
\end{itemize}

\xparagraph{Testbed.}
Our testbed consists of eight servers running Linux kernel 4.4.
Each server has two sockets with 14-core 2.2Ghz Intel Xeon Gold 5120 CPUs,
  192~GB of DRAM, two SSDs with 381GB and 1.8TB, connected to a
  10~Gbps network.
The client web browser runs on a laptop connected to the servers
  via a 100Mbps network with 1ms ping time to the servers.
This setup represents a typical enterprise setting.

\xparagraph{Dataset.}
We use a dataset with US airline flight performance metrics for the past
  20 years~\cite{ontime}.
Each row has a flight with its origin, destination,
  flight time, departure and arrival delays, etc.
This is a real dataset with numerical, categorical, text, and undefined
  values.
There are 130 million rows and 110 columns,
  which amount of 58.2~GB of uncompressed data.
In some experiments, we scale the dataset by a factor of 5, 10, or 100,
  by replicating its rows and reading them repeatedly from disk.
These datasets are labeled ``Flight-Kx'' where $K{=}1,5,10,100$ indicates
  the replication factor ($K{=}1$ is the original dataset).

\subsection{\sysname end-to-end performance}\label{sec:eval:macro}

We measure the end-to-end time that \sysname takes to execute spreadsheet
  operations for datasets of various sizes.

\xparagraph{Baseline.}
We compare \sysname against the traditional approach for big-data spreadsheets,
  such as Tableau, which is to connect a visualization front-end to a
  general-purpose analytics back-end.
Our baseline uses Spark as the back-end, and we measure only the
  analytics delay (not the rendering delay), giving an advantage to the baseline.
We optimize Spark to our best ability. 
We write queries in Scala; we pre-load all data to RAM before measuring; 
  and we use the same optimizations for each query as \sysname, including sampling.


\xparagraph{Workload.}
Figure~\ref{fig:spreadsheetops} shows the visualizations
  we are measuring.
We picked these operations using two criteria:
  (1) Each group of operations corresponds to a user action
      in the spreadsheet (\eg, ask for a histogram, or change the sort order of a tabular view).
  (2) The operations covers a broad range of the vizketches available in
      \sysname.

\begin{figure}
{\footnotesize
\begin{center}
\begin{tabular}{cl}
  \textbf{Name} & \textbf{Description} \\
  O1 & Sort, numerical data \\
  O2 & Sort 5 columns, numerical data \\
  O3 & Sort, string data \\
  O4 & Quantile + sort, 5 columns, numerical data \\
  O5 & Range + (histogram \& cdf), numerical data \\  
  O6 & Filter + range + (histogram \& cdf), numerical data \\
  O7 & Distinct + range + histogram, string data \\ 
  O8 & Heavy hitters sampling, string data \\
  O9 & Distinct count, numerical data \\
  O10 & Range + (stacked histogram \& cdf),numerical data \\
  O11 & Heatmap, numerical data \\
\end{tabular}
\end{center}
}
\vspace{-3mm}
\caption{Spreadsheet operations. The $+$ indicates serial
  operations, while $\&$ indicates concurrent operation.
  Numerical data refers to integer or floating point.}
\label{fig:spreadsheetops}
\end{figure}

\begin{figure*}
	\begin{center}
		\includegraphics[width=6in]{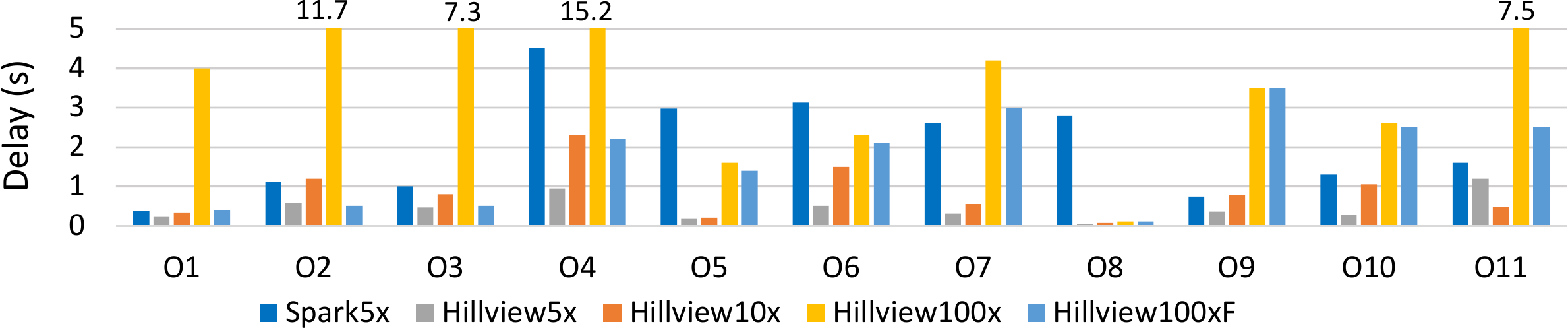}
		
		\includegraphics[width=6in]{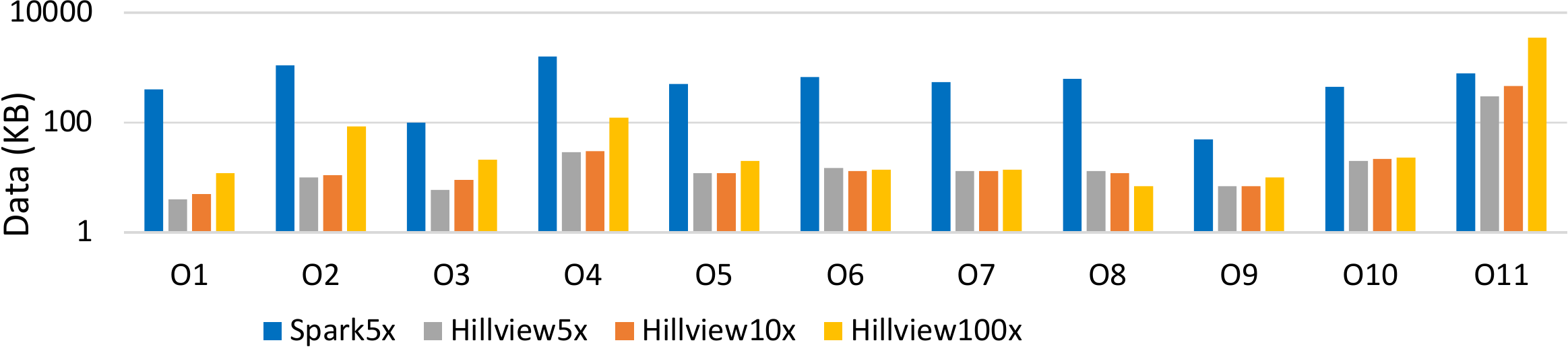}
	\end{center}
	\caption{End-to-end performance comparison.
		The top graph shows the response time to produce each visualization, while the bottom
		graph shows how many bytes the root node received.
		Here, we ensure the data is in memory before the measurement starts.
		The bars are labeled with the system name (Spark or \sysname) and the dataset size
		(5x to 100x corresponding to 650M to 13B rows).
		{Hillview100xF} is the time it takes for \sysname to produce the first
		partial visualization running with 100x
                data.
	}
	\label{fig:macro}
\end{figure*}

\begin{figure}
	\begin{center}
		\includegraphics[width=3in]{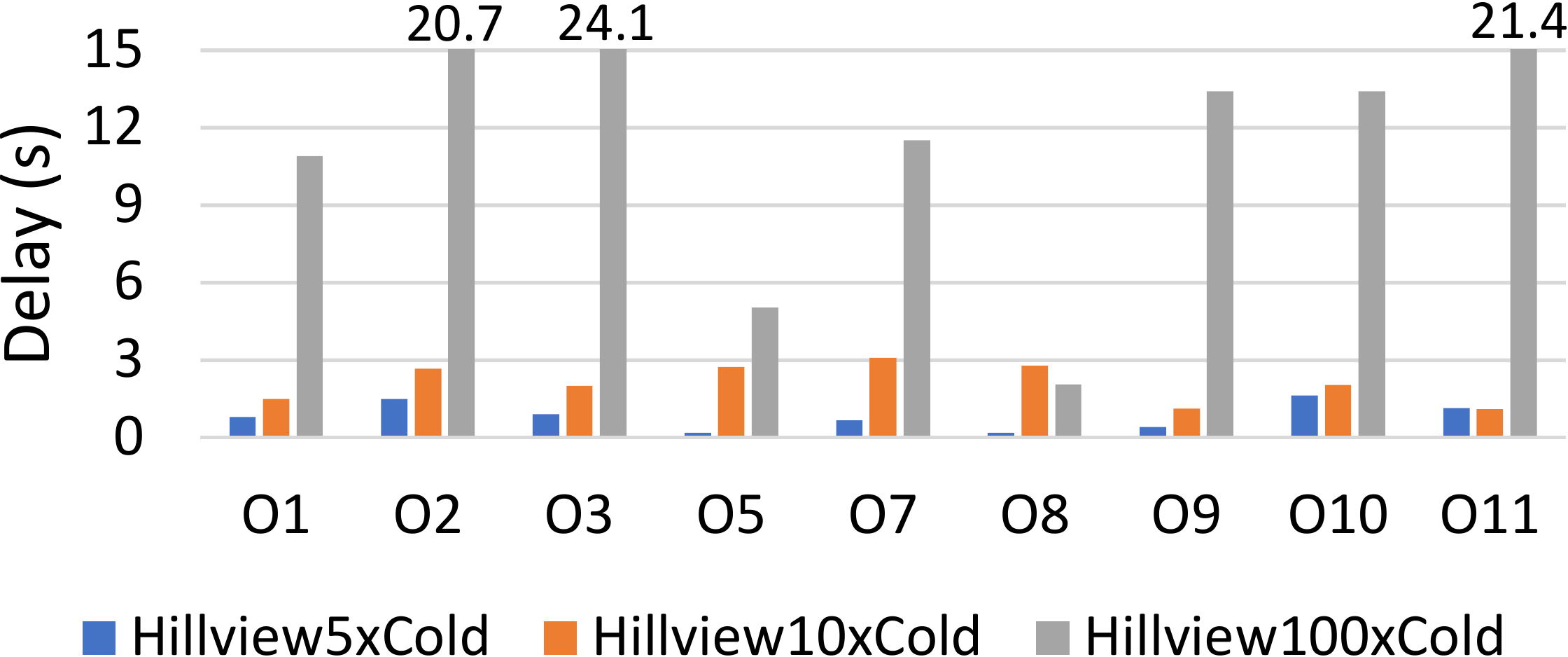}
	\end{center}
        \vspace{-4mm}
	\caption{End-to-end performance of \sysname when data is not in memory, so
		it needs to be loaded from SSD. Not shown are first visualizations, which
		arrive within 2.5s most of the time, and within 4s always.
                O4 and O6
		are omitted because in the spreadsheet these operations never happen
		with cold data (a prior action loads the data).
	}
	\label{fig:macro-cold}
\end{figure}

\xparagraph{Setup.}
In each experiment, we pick an operation, a dataset size, and a system.
The dataset sizes vary from 5x--100x the original data, corresponding to
  650M--13B rows of data with 110 columns each, for a total of
  71B--1.4T cells.
We submit the operation to the system and measure its response time
  and amount of data received by the root node.
For \sysname, we submit the operation at the user interface
  of the web browser, and we measure two response times at the browser:
  first partial visualization and final visualization.
For the Spark baseline, we start the
  measurement when the computation starts, and end the measurement
  when the query result is obtained.
For \sysname, we consider two cases: data is in memory
  before the measurement, and data is cold on disk (SSD).
For Spark, we only consider the case with data in memory.
 
\xparagraph{Results.}
Figure~\ref{fig:macro} shows the results for warm data in memory.
We could not run Spark with a dataset larger than 5x because it exhausted the memory at the servers: for example, the 10x dataset has 582~GB on-disk but its in-memory
  representation expands beyond the available aggregate memory in the testbed.

The top graph shows the response time.
We see that for most operations, \sysname performs at least as well as
  Spark, even when \sysname processes twice the data.
We also see that \sysname at 100x can be slow to compute all results: 7.3--15.2s.
However, \sysname produces a partial visualization quickly, which provides a
  better interactive experience.

The bottom graph shows the amount of data received over the network by the root node (for \sysname) or the
  master (for Spark); note that the Y axis is log-scale.
Spark consumes an order of magnitude more bandwidth than \sysname, except
  for O11.
This is because \sysname transmits a small amount of data to produce the visualizations.
The exception, O11, is a heatmap, which contains a large number of cells and hence
  its vizketch carries considerable more data. 
We also see that \sysname consumes more bandwidth with a larger dataset.
This is because the larger dataset takes longer to complete, and so \sysname
  transmits partial visualizations; with O11, the total amount of data
  becomes larger than Spark, but it is still reasonable at 3.5MB.
  
Figure~\ref{fig:macro-cold} show the results for cold data on disk.
For 5x and 10x data, visualizations still complete in 3s.
For 100x, the delay can be 24s; first visualizations
  arrive earlier, often within 2.5s (not shown).
  
In all cases, \sysname provides acceptable performance for interaction.
In our experience using \sysname, we tend to spend significantly more
  time browsing and analyzing charts than waiting
  for visualizations (cf \S\ref{sec:casestudy}).

\subsection{Vizketch microbenchmark}\label{sec:eval:micro}

We now consider the base performance of vizketches on one thread, and
its scalability over threads and servers.  We run each measurement
multiple times, and we display the variance of measurements after
excluding the fastest and slowest measurements; the variance tends to
be small\footnote{The first measurement warms up the Java JIT
  compiler, so it is generally much slower.}.

\xparagraph{Workload.}
We benchmark two types of histograms vizketches: one based on
  sampling (approximate, with bounded error) and the other based on streaming (no error).
We run these on numeric data.

\subsubsection{Single thread performance}\label{sec:evaldb}

\xparagraph{Baseline.}
The baseline is a common high-end commercial in-memory database system
  performing a histogram calculation; we are not allowed to reveal its name.

\xparagraph{Setup.}
In each experiment, we pick a computation method (streaming, sampling, or database system).
We measure the time it takes to execute the method
  on a single thread on 100 million rows.
For vizketches, we use a tree with a single leaf directly connected
  to the root, limiting execution to a single thread.
For the database system, we do not constrain the number of
  threads that it uses.

\xparagraph{Results.}
We obtain the following measurements:

\begin{center}
\begin{tabular}{lr}
\textbf{Method} & \textbf{Time (ms)} \\
streaming  & 527 \\
sampling & 197 \\
database system  & 5,830 \\
\end{tabular}
\end{center}

We see that the database system is an order of magnitude worse, because
  it has overheads that vizketches avoid: data structures must support
  indexes, transactions, integrity constraints, logging, queries of
  many types, etc. (although none of these are necessary in our case).
In contrast, vizketches are specialized to perform only the required computation.


\subsubsection{Scalability to multiple CPUs}

We now consider the performance of vizketches as we
  run them on multiple CPUs.

\xparagraph{Setup.}
We consider a computation tree that has $n$ leafs
  on the same server, connected to a single root.
The system executes each leaf on a separate thread,
  up to the available CPUs in the system.
In each experiment, we pick a number $n$.
As we increase $n$, we also increase the number of
  rows to be processed by adding more shards to the system,
  keeping constant the number of rows
  that each leaf gets---thus, the total number of leafs and the work
  increase together as $n$ grows.
We expect an approximately constant running time.

\begin{figure}
	\begin{center}
		\includegraphics[width=3in]{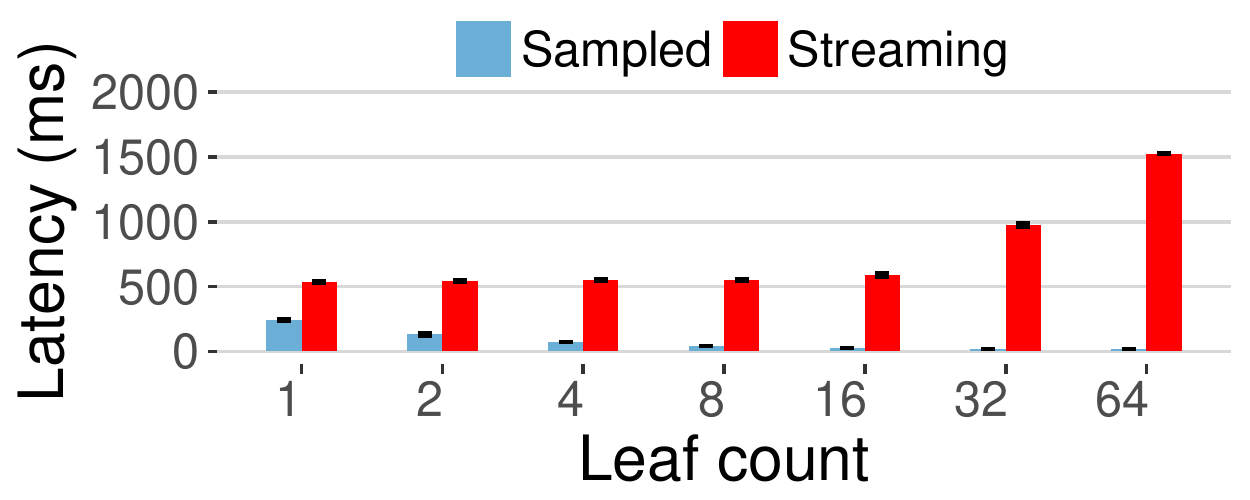}
	\end{center}
        \vspace{-3mm}
	\caption{Scalability of vizketches as we add more leafs and shards together.
		Ideal scalability would be constant latency.}
	\label{fig:scalethreads}
\end{figure}

\xparagraph{Results.}
Figure~\ref{fig:scalethreads} shows the results.
For the streaming histogram vizketch,
  we can see that latency remains constant up to 16 shards, showing
  a nearly ideal scalability up to that point.
After that, the server relies on hyper-threading, which
  impairs scalability.
For the (sampled) histogram vizketch, scalability is
  super-linear, because
  the sample size to obtain a given level of accuracy
  remains the same irrespective of the dataset size (\S\ref{sec:vizchart}).
Thus, as we add more leafs, we decrease the number of samples (and work done) per leaf.


\subsubsection{Scalability to multiple servers}

Next, we consider the performance of vizketches as
  we run them on many servers.


\xparagraph{Setup.}
In each experiment, we pick a number $n$ of servers and
  a vizketch.
We use a computation tree that has 64 leaf nodes on each
  server, connected to the root.
As we increase $n$, we increase the number of rows by
  adding more shards, so that each leaf node maintains
  the same number of shares (and rows).
We measure the time it takes to execute the vizketch
  running across the servers.

\begin{figure}
\begin{center}
  \includegraphics[width=3in]{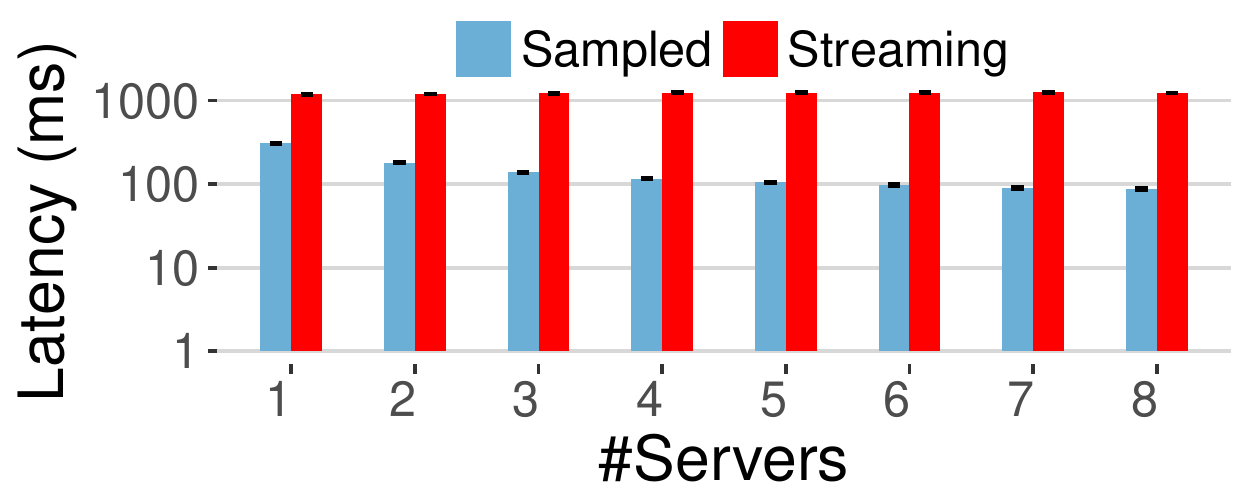}
\end{center}
\vspace{-5mm}
\caption{Scalability as we add more servers and increase the dataset proportionally.
		As before, the ideal scalability corresponds to a constant latency.  Note that the Y axis is logarithmic.}
\label{fig:scaleserver}
\end{figure}

\xparagraph{Results.}
Figure~\ref{fig:scaleserver} shows the result.
As before, for the streaming histogram vizketch, the latency remains constant
  as we add more servers and data, showing ideal scalability.
For the sampled histogram vizketch, we again observe
  super-linear scalability due to the same effect:
  the sample size remains constant, so the amount of work
  per server decreases with the number of servers.

\subsection{Vizketch applicability}\label{sec:eval:applic}

We consider our experience of using vizketch to
  implement the various spreadsheet functionality, to gain
  an understanding of the applicability of vizketches to
  processing data in \sysname.

When we started the project, we did not know if vizketches would suffice
  or we would need more powerful computation mechanisms.
In building the system, however, we found vizketches to be
  powerful and capable of implementing a broad range of
  functionality: tabular views, scrolling, simple data
  transformations, filtering,
  table summaries, and various visualizations.
We eventually realized that we could implement all data
  visualization functionality of \sysname using vizketches;
  in fact, \sysname has no other way to visualize data
  other than vizketches.

\subsection{Vizketch coding effort} \label{sec:eval:ease}

We now turn our attention to the effort required to write
  vizketches.
We again report on our experience with \sysname.

\begin{figure}
{ 
\begin{center}
\begin{tabular}{@{}lr@{\quad\quad\quad}lr@{}}
\textbf{Vizketch} & \textbf{LOC} & \textbf{Vizketch} & \textbf{LOC} \\
Histogram                 & 114          & Next items                & 191 \\
CDF                       & 114          & Find text                 & 108 \\
Stacked histogram         & 130          & Heavy hitters (sampling)  & 35 \\
Heatmap                   & 130          & Range                     & 156 \\
Heatmap trellis           & 127          & Number distinct           & 117 \\
Quantile                  & 79           & \\
\end{tabular}
\end{center}
}
\vspace{-3mm}
\caption{Effort required to implement vizketches.
}
\label{fig:loc}
\end{figure}

Quantitatively, Figure~\ref{fig:loc} shows the number of
  lines of back-end (Java) code required to implement each vizketch in \sysname.
We can see that the code is compact: the largest vizketch
  takes only 191 lines of code.
We found that an expert takes only a few hours to implement and test the code.
However, some vizketches involve fairly sophisticated algorithms; selecting or
  developing those algorithms
  took considerably longer than implementing them.
  In general, developing the UI to display the data and provide user
  interaction requires considerably more effort.

Qualitatively, implementing vizketches never required
  thinking about distributed systems or concurrency.
A developer simply provides the \t{summarize} and \t{merge} functions,
  which are purely local, while the rest of \sysname takes
  care of the distributed systems aspects of running
  vizketches across many cores and servers.
Of course, we had to implement the distributed
  execution framework for vizketches in \sysname, but
  this implementation was done once and benefits all
  vizketches, including future extensions.

\subsection{\sysname effectiveness: case study}\label{sec:casestudy}

We next consider the question of how effective
  \sysname is to browse and answer queries on large datasets.
We address this question through a case study.

\begin{figure}
\noindent
{
\begin{center}
	\begin{tabular}{@{}c@{\,\,}p{2.8in}@{}}
         \textbf{Question} & \textbf{Description} \\
         Q1 & Who has more late flights, UA or AA? \\
         Q2 & Which airline has the least departure time delay? \\
         Q3 & What is the typical delay of AA flight 11? \\
         Q4 & How many flights leave NY each day? \\
         Q5 & Is it better to fly from SFO to JFK or EWR? \\
         Q6 & How many destinations have direct flights from both SFO and SJC? \\
         Q7 & What is the best hour of the day to fly? \\
         Q8 & Which state has the worst departure delay? \\
         Q9 & Which airline has the most flight cancellations? \\
         Q10 & Which date had the most flights? \\
         Q11 & What is the longest flight in distance? \\
         Q12 & Is there a significant difference between taxi times of UA or AA on the same airport? \\
         Q13 & Which city has the best and worst weather delays? \\
         Q14 & Which airlines fly to Hawaii? \\
         Q15 & Which Hawaii airport has the best departure delays? \\
         Q16 & How many flights per day are there between LAX and SFO? \\
         Q17 & Which weekday has the least delay flying from ORD to EWR? \\
         Q18 & Which day in December has the most and least flights? \\
         Q19 & How many airlines stopped flying within the dataset period? \\
         Q20 & How many flights took off but never landed? \\
	\end{tabular}
\end{center}
}
\vspace{-3mm}
\caption{Questions used to evaluate the effectiveness of \sysname
  at extracting information from data.}
\label{fig:queries}
\end{figure}

\xparagraph{Workload.}  A person who is not familiar with \sysname
examines the Flights-1x data set and formulates a set of questions
(shown in Figure~\ref{fig:queries}) that interests her and that she
thinks the dataset answers.

\xparagraph{Setup.}
The experiment is carried out by an operator who is familiar with \sysname
  well but does not know the questions ahead of time.
In each experiment, we show a question to the operator
  and ask him to answer it using \sysname.
Our goal is to understand if the spreadsheet is powerful enough to
  answer the question and, if so, how easily it can do that.
Note that this experiment does not evaluate ease-of-use by beginners,
  because the operator is an expert.
This is intentional: \sysname users are not casual users but data analysts,
  whose job is dedicated to explore data and so they can obtain the required training.
  
For each question, we measure the time and number of spreadsheet actions that 
  the operator takes to answer the question.
A spreadsheet action consists of choosing an operation on a menu,
  clicking on the spreadsheet, or dragging the mouse to select a region.
For example Q1 can be answered by
  filtering the main table for column Airline=UA,
  producing a histogram on DepartureDelay, then going back to the
  main table and filtering for column Airline=AA,
  producing a second histogram on DepartureDelay.
To answer the question, we hover the mouse over the histograms
  to find the delay percentiles.
  
\begin{figure}
{ 
\begin{center}
\begin{tabular}{c@{\,\,}c@{\,\,}l@{\quad\quad\quad}  c@{\,\,}c@{\,\,}l}
\textbf{Question} & \textbf{Actions} & \textbf{Time}  & \textbf{Question} & \textbf{Actions} &  \textbf{Time}  \\
Q1 & 5 & 1:11   & Q11 & 3 & 1:18 \\
Q2 & 3 & 1:32   & Q12 & 5 & 6:44 \\
Q3 & 4 & 1:13   & Q13 & 6 & 6:27 \\
Q4 & 5 & 0:47$^{\ast}$ & Q14 & 2 & 0:20 \\
Q5 & 5 & 2:26  & Q15 & 4 & 1:56 \\
Q6 & 4 & 2:15$^{\ast}$  & Q16 & 3 & 1:07 \\
Q7 & 2 & 1:08  & Q17 & 3 & 1:07 \\
Q8 & 5 & 2:56  & Q18 & 2 & 1:08 \\
Q9 & 1 & 0:34  & Q19 & 2 & 0:40 \\
Q10 & 1 & 1:08$^{\ast}$ & Q20 & --- & 2:23$^{\dag}$ \\
\end{tabular}
\end{center}
}
\vspace{-4mm}
\caption{Number of actions and time in minutes:seconds required for an
  operator to answer questions using \sysname.  Most of the time is
  spent thinking about how to best translate a question into a set of
  UI operations.
Notes: 
$^{\ast}$These queries had only a partially satisfactory answer.
$^{\dag}$In this question, the data set did not have enough information to answer it;
   the measured time
   is how long it took to make that determination.}
\label{fig:queriesresult}
\end{figure}

\xparagraph{Results.}
Figure~\ref{fig:queriesresult} shows the results.
Answering a question took at most 6:44 (minutes:seconds), with most questions
  taking less than 2:30 (all except three).
The average and median times are 1:57 and 1:12.
Most of the time is the operator thinking about what to do,
  rather than waiting for the spreadsheet to respond
  (if the operator knew exactly what to do, all queries
  could be answered in under 30 seconds).
The minimum and maximum number of actions
  were 1 and 6, with mean and median 3.4 and 3.
Queries Q4, Q6 and Q10 did not have completely satisfactory
  answers because the spreadsheet cannot clearly
  separate dates (Q4, Q10) or the spreadsheet did not
  merge and deduplicate the destinations (Q6).
Question Q20 could not be answered because the dataset
  does not have the information (\eg, we discovered that
  it lacks the downed flights on 9/11).
We see that \sysname was effective at addressing
  most queries at small times, showing that
  (1) \sysname implements enough functionality to be usable and
  (2) it provides a interactive experience for human
  timescales.


%
%
%
%
%

\section{Related work}\label{sec:related}

\sysname is the first spreadsheet to scale massively
   with interactive speed.  \sysname
borrows ideas from the algorithms and computer graphics literature,
namely mergeable summaries~\cite{Agarwal2012} (or sketches) and
visualization-driven computation; it uses relies on many techniques
from databases (approximate query processing, on-line analytics),
big-data analytics, and distributed systems.

\sysname follows Shneiderman's visualization mantra~\cite{Shneiderman-svl96}:
``overview first, zoom and filter, details on demand''.
Fisher~\cite{Fisher-hilda16} identifies principles for 
  interactively visualizing big data (``look at less of it'' and ``look at it
faster''); these principles guided the design of
vizketches.


Big data visualization is a broad area;
  we give an overview of the closest related work below.
For more information, we refer the reader to several surveys in the 
  area~\cite{Scheidegger-bigdata16,godfrey-tkde16,Ghosh-vi2018,behrisch-tvcg19,Bikakis2018}.
Compared to published systems, \sysname achieves the best
  scalability for the amount of resources: we are not aware
  of any system that can handle a trillion cells with only 8 servers.

\xparagraph{Distributed visualization engines.}
\sysname evolved from Sketch~\cite{budiu-egpgv16}, which
  proposes a distributed data exploration library with applications to    
  a performance analyzer and a spreadsheet.
VisReduce~\cite{im-bigdata13} provides
incrementally updated approximations of visualizations computed over
progressively larger samples.  Vizdom~\cite{Crotty-vldb15} is a simple
UI for data manipulation and exploration; it runs on top of the A-WARE
smart caching and streaming engine~\cite{Crotty-hilda16} and uses the
Tuppleware analytics system~\cite{Crotty1-vldb15}.

\xparagraph{Visualization using big data query engines.}
One way to visualize big data is to connect a visualization
  engine to an analytics engine, such as
Hive~\cite{Thusoo-vldb09}, Impala~\cite{kornacker-cidr15},
Presto~\cite{presto18}, Dremel~\cite{Melnik-vldb10} (commercialized as
BigQuery), Drill~\cite{Apache-drill}, PowerDrill~\cite{hall-vldb12},
Spark~\cite{Zaharia-sosp13}, Druid~\cite{Yang-sigmod2014},
or Pinot~\cite{Im-sigmod18}.
This approach has advantages: it reduces design effort by using
  existing systems, and it leverages the years of effort spent in their
  optimization.
However, this approach does not achieve the
  speed needed for a spreadsheet: the generality of  
  analytics engine imposes overheads and computes unnecessary results,
  since there is no integration with the visualization engine.
Several systems follow this approach.  Microsoft
PowerBI~\cite{powerbi} using DirectQuery~\cite{jonge-white17} and
Polaris/Tableau~\cite{Stolte-cacm08,Wesley-sigmod11,Wesley-sigmod14}
provide plug-ins to many analytics engines; as discussed
in~\cite{christopher-tableau19}, the users of such systems have to
carefully avoid many queries that cannot be answered efficiently.  IBM
BigSheets~\cite{brown-bigsheets13} computes interactively only over a
subset of the data; once the user settles on a query, it is actually
run in batch mode using Spark.  HadoopVis~\cite{eldawy-icde16} uses
Hadoop to render geo-spatial data.  ScalaR~\cite{battle-bigdata13}
uses relational databases; the system in~\cite{vo-ldav11} uses
MapReduce for mesh rendering and isosurface extraction.
SwiftTuna~\cite{jo-vis16,jo-pacificvis17} uses Spark.
OmniSci~\cite{omnisci18} uses GPUs in one machine for server-side
rendering.
 
Facebook's Scuba~\cite{Abraham-vldb13} has been used as a back-end to
visualization systems.  Scuba provides fast response times but
with a different trade-off between data scale, responsiveness, and
correctness.  Scuba computes much more than ``what you can see'', since
its compute engine is decoupled from its visualization.  Thus, 
queries might return unbounded amounts of data to the visualization
engine, hampering real-time responsiveness.  To avoid that, Scuba
truncates worker responses to 100,000 rows (\cite[page~4]{Abraham-vldb13})
 and omits workers that do not respond in 10ms
(\cite[page~6]{Abraham-vldb13}). This can produce arbitrarily incorrect
visualizations.

The vizketch computational model is similar to the MPI
Reduce~\cite{MPI96} primitive used in supercomputing, to the Neptune
system~\cite{chu-ppopp03}, to the architecture of log analytics
systems such as Splunk~\cite{Splunk}, and to aggregation networks for
sensor networks~\cite{Rajagopalan-cst06}; these are general-purpose
platforms, and not visualization systems.

\xparagraph{Sampling and indexing.}
Sampling and indexing are used to accelerate visualization in many
systems.  \cite{Chaudhuri-sigmod01} considers the problem of sampling
a database for minimizing the error for a given set of queries.
BlinkDB~\cite{agarwal-eurosys13} uses stratified sampling, which is effective,
  but leaves the burden on users to write appropriate SQL queries
  and find appropriate error and time bounds.  Smart sampling
is used by~\cite{Fisher-chi12}.  \cite{wagner-sigplan16} uses
stratified sampling to accelerate queries in log management systems.
\cite{yan-vldb14} uses stratified sampling in Scope to reduce sample
sizes while minimizing errors; samples are incrementally maintained.
Pangloss~\cite{moritz-chi17} uses ``optimistic'' visualization on
sampled data to provide fast results.

The idea of using perceptual limitations to drive sampling appears
first in~\cite{Dix-avi02}.  \cite{Kim-vldb15} uses perceptual
limitations and sampling algorithms for specific chart types (e.g.,
bar charts).  M4~\cite{jugel-pvldb14} uses the screen resolution to
rewrite SQL queries to compute reduced results suitable for renderings
of line plots; this is extended for other chart types in
VDDA~\cite{Jugel-vldb16}.  Sample+Seek~\cite{ding-sigmod16} executes
responsive aggregated queries on a single table; it uses
measure-biased sampling together with new indexing schemes, specific
to the aggregation computed, to minimize errors.
G-OLA~\cite{zeng-sigmod15} handles interactive aggregate OLAP queries
over massive data sets.

VAS~\cite{park-icde16} samples data to minimize the visualization
errors for scatter-plots.  SynopViz~\cite{bikakis-sw17} and
Skydive~\cite{godfrey-edbt16} build hierarchical multi-scale models of
the data for browsing linked data sets.

\xparagraph{Progressive analytics.}
\sysname visualizations are incrementally updated; this technique is
called online aggregation~\cite{hellerstein-sigmod97} or progressive
analytics~\cite{Fisher-chi12,stolper-tvcg14,Turkay-tvcg17,fekete-arxiv16}.
There is significant work on this topic.
MapReduce
Online~\cite{Condie-nsdi10,pansare-pvldb11} is based on MapReduce.
EARL~\cite{Laptev-pvldb12} uses statistical boostrapping for providing
reliable on-line early estimates for the output of MapReduce
computations.  Progressive Insights~\cite{stolper-tvcg14} finds common
subsequences in event series of medical records, focusing on
  its UI design for incremental display and
  exploration.
 PIVE~\cite{Choo-vast14} adapts computation to limited
screen resolution for iterative algorithms (such as clustering or
dimensionality reduction).  DimXplorer\cite{Turkay-tvcg17} performs
progressive computation and rendering of dimensionality reduction
operations (such as clustering and PCA); it uses sampling for fast
response times.  Stat!~\cite{Barnett-sigmod13} operates in conjunction
with a streaming engine, and presents immediately incremental results.
Microsoft's Tempe~\cite{Tempe} runs on top of a streaming engine and
provides progressive visualizations.

All above systems lack some of the benefits of vizketches:
parallelization, computation efficiency, and bandwidth efficiency
(\S\ref{sec:benefits}), which are required for \sysname.

Nanocubes~\cite{lins-tvcg13}, imMens~\cite{liu-eurovis13}, and
Hashedcubes~\cite{cicero-tvcg17} improve interactivity by
pre-processing the data to build smart indexes.
DICE~\cite{kamat-icde14}, Sesame~\cite{Kamat-kdd18} and
ForeCache~\cite{battle-sigmod16} use sessions and locality to
pre-compute views or to reuse computation results between consecutive
views.  These ideas improve interactivity, but restrict the scope of queries to
pre-processed columns (\eg, the user pre-selects a few columns to optimize) or spend
significant time for pre-processing. By contrast, \sysname uses no pre-processing
or indexes, because we do not know ahead of time which columns the
user might choose to explore.

VisTrees are indexes designed to support quick histogram construction
for visualizations~\cite{elhidni-hilda2016}.
Profiler~\cite{kandel-avi12} and Foresight~\cite{Demiralp-VLDB17}
propose methods to find abnormality in the data; \sysname could
incorporate this functionality, especially Foresight which is based on
sketches.  NeedleTail~\cite{Kim-hilda18} uses a small in-memory index
to allow fast browsing and displaying any-k records.
AQP++~\cite{Peng-sigmod18} combines approximate query processing with
aggregate pre-computation.

\section{Conclusion}

\sysname is a spreadsheet that supports
  a trillion cells even with a modest number of servers.
\sysname introduces a new query execution engine specialized
  to render tabular views and charts for a spreadsheet.
The new engine uses vizketches, a new
  but simple idea that parallelizes computation and calculates only
  what is needed for a good visualization.
We believe \sysname is a useful tool for humans to
  explore data; it nicely complements other tools, such as
  analytics frameworks, which have other uses.

\vfill

\nocite{Cohen07}
\nocite{dean-osdi04}
\nocite{Feldman-TA10}
\nocite{FlajoletFGM07}
\nocite{misra-scp82}
\nocite{Muthukrishnan2005}
\nocite{RubinfeldS2011}
\nocite{MLbook}
\nocite{thorup13}

\bibliography{paper}
\bibliographystyle{abbrv}
\if\confversion0
\appendix
\section{A Computational model for vizketches}\label{sec:model}

\begin{figure}[t]
  \centering
  \includegraphics[width=.5\columnwidth,trim=0in 3.6in 10.3in 0in]{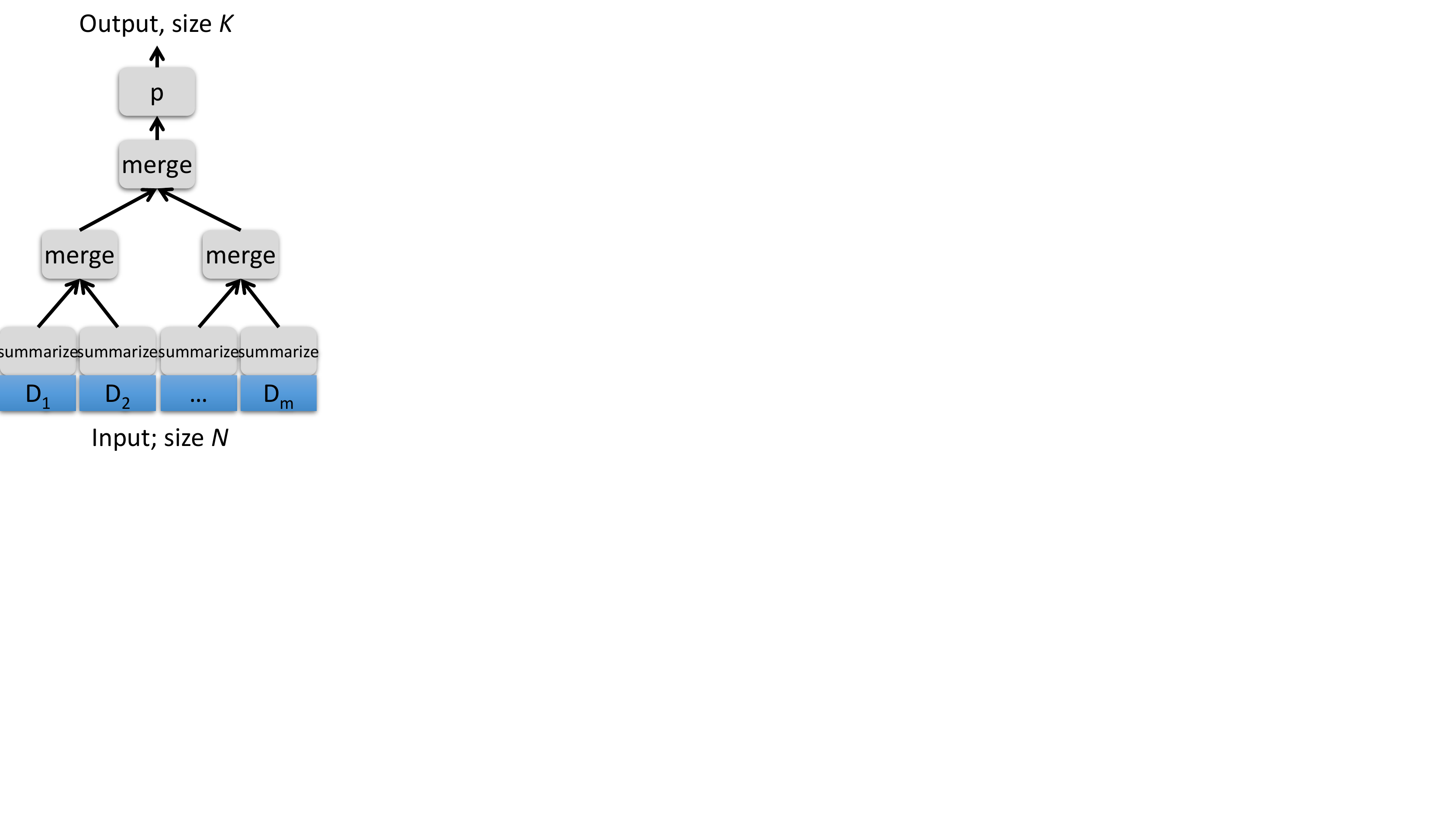}
  \caption{Abstract computational model for vizketches.}
  \label{fig:model}
\end{figure}

\newcommand{\V}{\ensuremath\mathcal{V}}

The abstract computational model for vizketches is illustrated in
Figure~\ref{fig:model}: consider a rooted tree $T$ with $m$
leaves\footnote{The tree does not have to be balanced as in this
  figure.}  $\{\ell_1,\ldots, \ell_m\}$.  Each leaf is a server that $\ell_i$ contains
a local dataset $D_i$, which is usually a multiset of values from a
domain $\D$. The various $\ell_i$s can run computations in parallel.
Let $D =\uplus_{i=1}^nD_i$ denote the entire input which is also a
multiset over $\D$ (where $\uplus$ denotes disjoint multiset union).
We will denote with $N = |D|$ the size of the input data. These same
servers can also servers as internal nodes (and root) of the tree.

Let $\V$ be the space of data views (histograms, for instance).
There is a client who does not know the input, and who wishes to
compute a function $f(\D) \in \V$, where each element of $\V$ is $v$
bits long, and is essentially {\em independent
of the size of the input $N$}.  We stress that $f$ is a function on
multisets from $\D$, and not on sequences, so it is oblivious to how the records are
partitioned among the servers and how each server orders them.

A computation protocol proceeds in $k$ rounds for some integer $k$, in all the
vizketches we implement $k$ was $1$ or $2$.  In round $j$ where $j \in
[k]$, the root issues a request $r^j$, which is broadcast to all
servers.  The request is based on the function $f$ and the results of
previous rounds. The computation then proceeds in two phases:

\begin{itemize}

\item {\bf Summarize: } Every leaf makes a pass over their respective input and computes a function
        $\crt^j: D_i \rgta \V$ and sends it to its parent
        in $T$. While each round could do something different
        the description length of elements output is
        bounded by $v$ for every round $j$.

\item {\bf Merge: } This phase applies an aggregation function $\agg^j:
  \V^d \rgta \V$ at each internal node where $d$ is
  the number of children. The function aggregates results from
  the children and sends the aggregated
  results to the parent of $v$ in $T$.  The inputs to $\agg$ are
  from the same small domain $\V$, so the computational cost of $\agg$
  is dominated by the cost of the creation function $\crt$.
\end{itemize}

At the end of the last round, the root  computes the output $\hat{f}(D)$.
Protocols may be randomized. Approximation is often essential since computing even simple functions
exactly can be hard in this model. There are two requirements from the protocol:
\begin{itemize}
		\item {\bf Correctness: } $\hat{f}(D)$ approximates $f(D)$, under a
		suitably defined notion of approximation (with high probability for a randomized protocol).
		\item {\bf Efficiency: } The length of all messages communicated
		in the protocol and the memory needed to compute them are
		$\mathrm{polylog}(v, \log(n))$.
\end{itemize}

This model is closely related to well-studied models in the fields of
algorithms and databases. The $\crt$ phase relies on sampling
and streaming algorithms~\cite{RubinfeldS2011, Muthukrishnan2005}.
The $\agg$ phase and the model for incremental computations
is closely related to the sketching model and massive unordered data
(MUD) model~\cite{cormode-cacm17,   Feldman-TA10} of distributed
computation and the notion of mergeable
summaries \cite{Agarwal2012}.  
At the same time, the combination of the restriction on
the size of outputs in $\V$ and the efficiency requirement for \crt
are specific to visualization.  This implies that our
computational model is \emph{less} general than
map-reduce~\cite{dean-osdi04} where the communication between machines
can have messages of unbounded size.

\section{Vizketch algorithms}\label{sec:viz}

We present details for all classes of vizketches used in \sysname.

\subsection{Vizketches for charts}\label{sec:vizchart}

We describe the vizketches used in \sysname for producing various charts.
Here, a vizketch is parameterized by the target display resolution, and
  produces calculations that are just precise enough to render at that resolution.
We now give the details for each vizketch.
In appendix~\ref{sec:sample}, we give rigorous guarantees for correctness of the underlying algorithms

\begin{figure*}
\begin{tabular}{@{}c@{\hskip 10mm}c@{\hskip 10mm}c@{\hskip 10mm}c@{}}
\includegraphics{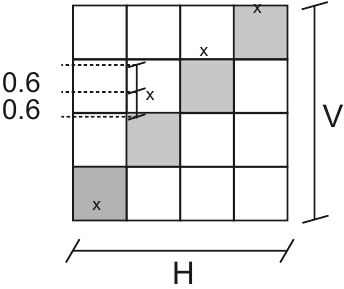} &
\includegraphics{figs/viz-hist} &
\includegraphics{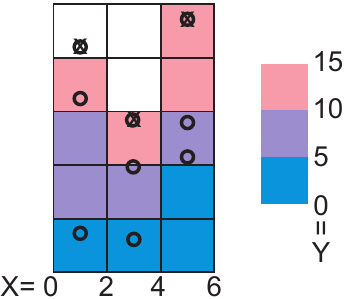} &
\includegraphics{figs/viz-heatmap}
\\
(a) & (b) & (c) & (d) \\
\end{tabular}
\caption{Charts in \sysname have an error of at most one pixel or one color shade
	     with high probability.
	(a) A cdf plot with dimension $H \times V$ pixels. The ${\times}$ in the figure indicates the exact value
	    of the cdf. The grey box indicates the rendering, which is at most 0.5
	    pixel away from the ${\times}$. 
	(b) a histogram with three bars. The ${\times}$ indicates the correct height for the bar---again,
	    at most one pixel away from the rendering.
	(c) A stacked histogram with three bars. The bars show a histogram for the first variable $X$.
	    Each bar has three subdivisions,
	    each with a different color, which indicates the value of the
	    second variable $Y$. The size of the subdivision shows the density for the bin with both
	    variables. The ${\circ}$ indicates the correct location for the subdivision.
	(d) A heat map. The x-axis has bins for the first variable; the y-axis, for the second variable.
	    The color indicates the density of each bin, where the error is at most one color shade
	    with high probability.
}
\label{fig:viz}
\end{figure*}

  \xparagraph{Equi-width buckets for string data.}
\sysname can plot charts for arbitrary string data.
It divides the data range into equi-width bins.
If there are few distinct values (50 or fewer), we assign a bin
for each value.
Otherwise, we consider an alphabetical ordering of strings
and aggregate contiguous values into bins, so that the number of
bins does not exceed 50.
The challenge is to find bin boundaries without sorting the full dataset.
This is done using a sketch based on bottom-k sampling~\cite{thorup13,Cohen07}, which
is an efficient mergeable randomized streaming algorithm
that computes \emph{approximate quantiles} over \emph{distinct}
strings, where the quantiles are set to 1/50, 2/50, etc.

\xparagraph{Cumulative distribution functions (CDFs).}
We are given a column with numeric values in
  a range $[x_0,x_1)$, and a target screen dimension of
    $H \times V$. Horizontally, a pixel $h$ represents an interval $I_h = [x_0, x_0 {+} (h{+}1)(x_1 {-} x_0)/H)$.
   We must determine how many data points belongs to $I_h$
  and divide by the total number of points.
   Even if we compute the cdf exactly, when we render it on the screen, we round it to the nearest pixel, which introduces a {\em quantization error} of ${\pm} 0.5/V$. A simple way to compute the cdf exactly is to scan the dataset and count, however, it requires a full scan and many counters.
%
In our renderings,  our goal is to compute, for each horizontal pixel of the cdf, a value between 0 and 1
  that lies within ${\pm}0.1/V$ of the true cdf value, so that after rounding,  we render a pixel value
  that is at most ${\pm} 0.6/V$ away from the true cdf value (and the pixel itself is at most one away for the pixel that would be rendered if we did an exact computation (Figure~\ref{fig:viz}(a)). The advantage now is that we can approximate the cdf to the desired accuracy level with probability $1 - \delta$ by just sampling the data.
Our calculation for the required sample size are based on a standard Chernoff bound for the error per pixel. We then use the fact that pixels are arranged on a line. In~\cite{extended} Appendix C we show the target sample size is $n=O(V^2 \log(1/\delta))$.
The \t{summarize} function thus samples the dataset with the appropriate rate and counts the
  number of values that fall in each interval $I_h$.
The function outputs a small vector $S$ of $H$ bin counts.
The \t{merge} function of the vizketch takes two such vectors and adds them.
To produce the cdf visualization from the vizketch, we scale each bin of $S$ by $V/n$
  to obtain a value in $0\ldots V{-}1$.

\xparagraph{CDFs for string data.}  We can produce CDF plots even for
string data, by combining the equi-width bucket computation with the
counting-based CDF computation described in the previous paragraph.

\xparagraph{Histograms.}
We are given a numerical column (or a value that can be readily converted to a real number, such as a date)
with range $[x_0,x_1)$, a number $B$ of histogram bars, and
their maximum pixel height $V$.
The histogram vizketch (Figure~\ref{fig:viz}(b)) is similar to the cdf vizketch, except that it has
$B$ bins instead of $H$ horizontal pixels and we now divide the range $[x_0, x_1)$ into $H$ equi-sized intervals, one per bin.
To maximize use of screen, we should scale the bars so that the largest one
  has $V$ pixels.
We show in Appendix C of~\cite{extended} that to bound the error of a bar to one pixel with probability $1-\delta$,
   requires sampling a set of size $n=O(V^2 B^2 \log(1/\delta))$.
The \t{summarize} function outputs a vector of $B$ bin counts, and the
  \t{merge} function adds two vectors.

\xparagraph{Histograms for text data.}  We allow users to plot
histograms for arbitrary text data; the number of bars is limited to
50, and if the number of distinct strings is larger each bar
represents a set of strings that are contiguous in alphabetical order.
The user can zoom-in to reveal the finer-grained structure of each
bar.  These are based on the equi-width buckets computation.

\def\ncolors{20}

\xparagraph{Stacked histogram.}
A stacked histogram (Figure~\ref{fig:viz}(c)) involves two columns $X$ and $Y$, which can be numeric or categorical.
We are given two input ranges $[x_0,x_1)$ and $[y_0,y_1)$, a
  number $B_x$ of histogram bars for $X$, a number $B_y$ of colors for bins for $Y$,
  and pixel dimensions $H \times V$.
The human eye cannot distinguish many colors reliably, so $B_y$ is limited to
  ${\approx}$\ncolors.
The stacked histogram represents counts in two ways: (1) the height of each
  histogram bar represents counts of bins of $X$ (like a histogram),
  (2) the height of a subdivision of a bar represents counts of a bin of $Y$
      within the bin of $X$ of that bar.
To bound the error by a pixel, the accuracy of (1) should be
  ${\pm}0.6$.
As for (2), in the worst case a subdivision will be the entire bar,
  which also requires an accuracy of ${\pm}0.6$.
The required target sample size is $n=O(V^2 B_x^2 \log(1/\delta))$.
The \t{summarize} function thus samples the dataset with the required rate,
  counting the number of values that fall in each $X$
  bin, as well as the number of values that fall in the combined $X,Y$ bins.
The function outputs a small vector $S$ of $B_x + B_x \times B_y$ bin counts.
The \t{merge} function of the vizketch takes two such vectors and adds them.

\xparagraph{Histogram (streaming).}
We also provide a simple vizketch to compute histograms
  without sampling, if users want to get the results precise to the last digit.
Given a numerical column with range $[x_0,x_1)$, a number $B$ of histogram bars,
  the \t{summarize} function scans the data and counts the number of items in
  each bin, producing a vector.
The \t{merge} function adds two vectors.

\xparagraph{Normalized stacked histogram.}
This visualization is like a stack histogram except that
  the size of each bar is normalized to 1.
This difference, however, impacts the required accuracy for subdivisions; for example,
  if a bin of $X$ has a small count, it gets normalized to a full bar and therefore its subdivisions
  require a higher accuracy.
To implement this visualization, we use a stacked histogram sketch without sampling.

\xparagraph{Heat map.}
We are given two columns $X$ and $Y$
  with ranges $[x_0,x_1)$ and $[y_0,y_1)$, and the pixel dimensions $H \times V$.
A heat map (Figure~\ref{fig:viz}(d)) defines bins in two dimensions, where each bin consumes $b \times $ pixels, where $b=3$.
Thus, we have $B_x = H/b$ and $B_y = V/b$ bins for $X$ and $Y$.
The density of a bin is represented by a color scale.  If we use $c{\approx}20$ distinct colors,
the required accuracy for each bin density is $1/2c$.
This requires a target sample size $n=O(c^2 B_x^2 B_y^2 \log(1/\delta))$ (sampling can only be used
if the mapping from count to color is linear, otherwise the full dataset has to be scanned).
The \t{summarize} function samples data with the target rate, counting the number of values
  that fall in each bin.
It outputs a matrix of $B_x \times B_y$ bin counts.
The \t{merge} function adds two such matrices.

\xparagraph{Trellis plots.}
A Trellis plot displays a 1D or 2D array of other plots; in Figure~\ref{fig:viz} we show a Trellis plots of heatmaps.
For a 1D Trellis plot are given three columns $W$, $X$, and $Y$; $k$ elements $w_1,\ldots,w_k$ of $W$;
  the ranges $[x_0,x_1)$ and $[y_0,y_1)$ of $X$ and $Y$; and pixel dimensions $H \times V$.
A heat map trellis plot
  produces $k$ heat maps, each for a fixed range of values $w_i$ in column $W$.
This might appear like significant computation, but because the rendering area
  is limited to $H \times V$, a large number of heat maps means that each
  heat map is small.
For example, if we render the $k$ heat maps as a $2 \times k/2$ matrix of heat maps,
  then each heat map has dimension $H/2 \times 2V/k$.
The vizketch computes all heat maps in parallel, but due to the quadratic dependency
  on the number of bins, this requires a smaller
  sample size than rendering a single heat map of the same pixel dimensions.
The \t{summarize} function outputs the same number of bins as a single heat map of the
  same pixel dimensions.

\subsection{Vizketches for tabular view}\label{sec:viztab}

The next several vizketches are used to produce the tabular views of the spreadsheet.
They follow the principle of calculating just what needs to be displayed,
  with approximations where possible.

\xparagraph{Next items.}
This vizketch is used to render a view of the spreadsheet given the current top row $R$
  (or $R=\bot$ to choose the beginning).
We are also given a column sort order, and the number $K$ of rows to show.
This vizketch returns the contents of the $K$ rows that follow $R$ in the sort order.
The \t{summarize} function scans the dataset and keeps a priority heap with the $K$
next values following row $R$ in the sort order.
The \t{merge} function combines the two priority heaps by selecting the
smallest $K$ elements and dropping the rest.

\xparagraph{Quantile for scroll bar.}
When a user moves the scroll bar she indicates a quantile of the data to be displayed.  The tabular will start display
   the appropriate quantile of the current sorting order corresponding to the scroll bar position.
For example, if the scroll bar is in the middle, we move to the median
  of the sorting column.
Given a scroll with $V$ pixels and a position $v$ in $0\ldots V{-}1$,
  this vizketch uses
  a quantile sketch method with target quantile $v/V$
  and accuracy
  ${\pm}0.6$, which samples $O(V^2)$ random rows and returns the quantile from this set.
The method returns the contents of a row that becomes the top entry
  in the tabular view.

\xparagraph{Find text.}
This vizketch implements the free-form text find functionality of the spreadsheet.
Given a row $R$, a search criteria (the search text; whether it is exact match, substring,
  or regexp; and whether it is case sensitive), and a column sort order,
  we want to find the next row satisfying the criteria in the sort order.
This is similar to the next item vizketch above except that we eliminate all rows that do not
  match the search criteria.

\xparagraph{Heavy hitters (streaming).}
This vizketch serves to find the most
  frequent elements of a column and their count.
More precisely, given a column and a maximum number $K$ of items, we want to find elements that
  occur more than a fraction $1/K$ of the time.
We also want the approximate counts for such elements.
To do this, we directly use the Misra-Gries
  streaming algorithm~\cite{misra-scp82}.

\xparagraph{Heavy hitters (sampling).}
Another vizketch to find heavy hitters works by sampling.
Given a probability of error $\delta$
  and let $K$ be as above.
The basic idea is to sample with a target size $n$ (determined below),
  and select an item as a heavy hitter if it occurs with frequency at least $3n/4K$.
A statistical calculation shows that by picking $n=K^2 \log(K/\delta)$,
  with probability $1-\delta$
  we can obtain all elements that occur more than $1/K$ of the time and no
  elements that occur fewer than $1/4K$ of the time.
This method is particularly efficient if $K$ is small.
In fact, our experiments indicate this method is better the previous one
 when $K \ge 1/100$.

\subsection{Vizketches for auxiliary functionality}\label{sec:auxviz}

\sysname uses vizketches to obtain general information about columns.
Technically, these vizketches are just sketches because they need not be
  tuned for a target visualization.
We list them here because in \sysname they are executed by the same
  mechanism as the other vizketches.

\xparagraph{Moments.}  Given a column, this vizketch collects its
minimum and maximum values, number of rows, the number of missing
values, and the statistical moments up to a specified value $K$
(including mean and variance, the first two moments).  This
information is used to select the range of certain visualizations
(histograms, heat maps, etc).  The information is also shown to the
user if she requests a summary of the column data.

\xparagraph{Number of distinct elements.}
This information is computed approximatively using the HyperLogLog sketch~\cite{FlajoletFGM07}.

\xparagraph{Principal component analysis.}
PCA can summarize $M$ numeric columns into $K {<} M$
columns, by projecting the $M {\times} N$ matrix into a $K {\times} N$
matrix along the eigen vectors of the $M {\times} M$ correlation matrix.
This matrix can be efficiently computed by a sampling-based sketch.

\subsection{Vizketches in the spreadsheet}

\begin{figure}
\begin{center}
	{\small
		\begin{tabular}{@{}p{1.2in}@{\hskip 2mm}p{1.5in}@{}}
			{\bf Spreadsheet functionality} & {\bf Required vizketch(es)} \\
			Drawing any chart               & Range + chart-specific vizketch \\
			Initial tabular view            & Next items \\
			Scroll up/down table            & Next items \\
			Moving scrollbar                & Quantile + next items \\
			Find text                       & Filter + next items \\
			Heavy hitters                   & Heavy hitters \\
                        PCA                                  & PCA \\
		\end{tabular}
	}
	\caption{Using vizketches to implement specific spreadsheet functionalities.}
	\label{fig:vizketchspreadsheet}
\end{center}
\end{figure}

Spreadsheet actions are implemented using one or more vizketches
  (Figure~\ref{fig:vizketchspreadsheet}).
All charts, when produced initially, require a vizketch to
  determine the range of the inputs; subsequently, this information can be
  cached.
Changing the tabular view requires the next items vizketch to produce
  the new visible rows sorted the chosen order.
The scrollbar is implemented by a quantile computation (\eg, scrollbar
  in the middle corresponds to the median), following by next items to
  produce the new visible rows.
Specific analyses run the corresponding vizketches (\eg, Heavy hitters).

\section{Proofs of Correctness}
\label{sec:sample}

Consider a distribution $\P$ on some domain $\D$ which might be
discrete or continuous. Let $\F$ be a family of subsets of $\D$. For $I
\in \F$, let $\mu_\P(I)$ be its measure under $\P$. We draw a sample $S$ of $n$ of
size $n$ from $\D$ according to $\P$.
We expect $|S \cap I|/|S|$, the fraction of points in $S$ that
land in $S$, to be close to $\mu_\P(I)$ for every set $I$.
The VC dimension theorem relates the sample size needed for this to the
VC-dimension (see \cite[Chapter 6]{MLbook}) of the family of subsets.

\begin{theorem}
  \cite[Theorem 6.8]{MLbook}
\label{thm:vc}
Let $\P$ be a distribution on $\D$, and let $\F \subseteq 2^{\D}$ be a
family of subsets with VC-dimension $d$. Let $S$ be a (multi)-set
of samples drawn from $\P$ where  $|S| \geq C(d +
\log(1/\delta))/\epsilon^2$. Then with probability $ 1- \delta$, for
every set $I \in \F$, we have
\begin{align}
  \label{eq:sampling}
  \left|\frac{|S \cap I|}{|S|} - \mu_\P(I)\right| \leq \delta.
\end{align}
\end{theorem}

We will use this theorem to derive sample complexity bounds for
various UI operations. These bounds guide our algorithms when setting up the sample size.

\subsection{Quantile Estimation}\label{sec:quantile}

Assume that we have $V$ vertical pixels in our display, and $N$ rows in the
table. Assume there is a sorted order on these rows, and define the
(relative) rank of the $i^{th}$ row to be $i/N$. Let $D =\{i/N\}_{i=1}^N$ denote the set of rows (identified by rank).
Assume that the user moves the scroll bar to pixel $j$. If we were
computing quantiles exactly,  scrolling to pixel $j$ would render the
screen whose top row has rank $j/V$. We relax this notion and view
pixel $j \in [V]$ as representing a set of rows, whose rank lies in
the interval $I(j) = (j/V - \epsilon, j/V + \epsilon)$. A valid output
  for our quantile estimation routine is to return any element from this range.

Our algorithm is to take a sample $S$ of uniformly random rows from $[V]$ an return the element
whose relative rank is closest to $j/V$.

\begin{theorem}
\label{thm:quantile}
  If $|S| \geq O(\epsilon^{-2}\log(1/\delta))$, the above algorithm
  returns an element from $I(j)$  with probability $1 - \delta$.
\end{theorem}
\begin{Proof}
  Let $\P$ denote the uniform distribution on $D$. Let $\F$ denote the
  class of intervals $(a, b] \subseteq [0,1]$, this class has
    VC-dimension $2$. Note that $\mu_\P(a,b] = b -a \pm O(1/N)$. We
      ignore the $1/N$ factor, since our assumption is that $N \gg V$.

To ensure that our algorithm returns a row from $I(j)$, it suffices that both the
intervals $L =[0,j/V - \epsilon]$ and $R = [j/V + \epsilon, 1]$ adjoining $I(j)$ satisfy
\eqref{eq:sampling} for $\epsilon/2$. If this is the case, then
\[ \frac{|S \cap L|}{|S|} \leq j/V - \epsilon/2,  \frac{|S \cap R|}{|S|}
    \leq 1 - j/V - \epsilon/2 \]
This means that all the elements whose relative rank in $S$ lie in
the range $(j/V - \epsilon/2, j/V + \epsilon/2)$ must come from $(L \cup R)^c = I(j)$
The sample complexity follows by Theorem \ref{thm:vc}.
\end{Proof}

In practice, we choose $\epsilon = 1/(2V)$ so that the intervals $I(j)$
give a disjoint partition of $[0,1]$, which requires sample complexity
$O(V^2)$ for constant probability of success.

\subsection{Histograms and HeatMaps}
\label{sec:histograms-heatmaps}

Assume we have a table containing a set of rows $D$ from a range
$\D$ which is ordered. By mapping each element of $D$ to its relative rank,
we can identify $D$ with a subset of points in $[0,1]$. The
uniform distribution on $D$ gives a distribution $\P$
on $[0,1]$. To draw a histogram, we divide $[0,1]$ into $B$ equal
intervals/buckets. Assume we have $V$ vertical pixels in our
histogram. For each $b \in [B]$, let $p(b)$ denote the fraction of the
population in the $b^{th}$ bucket and let $p_{\max}$ denote the largest value.

A natural choice for the vertical range is $[0,p_{\max}]$, so that a
bar reaching pixel $j$ represents probability mass of $j\cdot
p_{\max}/V$. If we knew each probability $p(b)$ exactly, we would snap
that bar to the closest pixel, so that pixel $j$ represents
the interval $I_j = [(j -  0.5)p_{\max}/V, (j + 0.5)p_{\max}/V)$. Let
us refer to this as the {\em ideal} histogram. Even in the ideal
histogram, there is a {\em rounding error} of $p_{\max}/V$ in the
rendering of $p(b)$: if the bar for bucket $b$ has height $j$, then we
know that $p(b) \in I(j)$ which has width $p_{\max}/V$.

We define an approximate histogram as one where pixel $j$ represents
probabilities from a  slightly larger range. Concretely in a
$\mu$-approximate histogram, every bar with height $j$ represents a
probability in the range $\widehat{I}_j = [(j -  0.5 - \mu)p_{\max}/V,
  (j + 0.5 + \mu)p_{\max}/V]$.  As long as $\mu < 0.5$, this ensures
that every bar is within a pixel of its height in the ideal
histogram.

Our algorithm for computing approximate histograms is the obvious one: we
take a sample of $n$  uniformly random rows from $[N]$ and use this to
determine empirical probabilities $\hat{p}(b)$ for each bucket $b \in
B$. We determine $\hat{p}_{\max}$ and then assign a bar of height $j$
to bucket $b$ where $j\hat{p}_{\max}/V$ is closest to $\hat{p}(b)$.

\begin{theorem}
\label{thm:histogram}
  With $n = O(V^2/(\mu p_{\max})^2\log(1/\delta))$ samples, the above algorithm
  computes a $\mu$-approximate histogram with probability $1 - \delta$.
\end{theorem}
\begin{Proof}
As in Theorem \ref{thm:quantile}, we identify $D$ with a discrete
subset of $[0,1]$, and each bucket with an interval.  We take
$\epsilon = \mu p_{\max}/2V$ and compute the sample complexity by
Theorem \ref{thm:vc}. With probability $1 - \delta$, we have
$|\hat{p}_{\max} - p_{\max}| \leq \epsilon$. If bucket $b$ maps to
pixel $j$, it implies that
\[ \hat{p}(b) \in [(j -  0.5)\hat{p}_{\max}/V,
  (j + 0.5)\hat{p}_{\max}/V]\]
hence $p(b)$ lies in the interval
\[ [(j - 0.5  -\mu/2)\hat{p}_{\max}/V, (j + 0.5 +
  \mu/2)\hat{p}_{\max}/V] \subseteq \widehat{I}_j\]
so the claim follows.
\end{Proof}

If we set $\mu$ and $\delta$ to some constants (say $0.1$ and $0.01$),
the sample complexity by Theorem \ref{thm:vc} is $O(V^2/p_{\max}^2)$. In
the worst case, $p_{\max}$ might be as small as $1/B$ (where $B$ is the
number of buckets), although typically it is much larger. In practice, we
have found that using $CV^2$ samples for constant $C$ works well.

A closely related problem is that of computing the CDF of the
distribution of values in a table. We think of each pixel
representing a single bucket, where buckets are obtained by dividing
the range into equal parts. But rather than the individual mass $p(b)$,
we want to plot the cumulative sum $\sum_{i \leq b} p(b)$. To map
this to the VC dimension setting, we identify bucket/pixel $b$ with
the interval $[0,b/V]$. Setting the accuracy to $1/2V$ is sufficient,
since the vertical range is $[0,1]$. So by a similar argument to
Theorem \ref{thm:histogram} above, $n  = O(V^2/\log(1/\delta))$
samples suffice.

Similar bounds can be derived for 2-d histograms. In this case,
there are sort-orders on two columns, which lets us map rows to a
discrete subset of $[0,1] \times [0,1]$. Each bucket is an
axis-aligned rectangle in $[0,1] \times [0,1]$, a class which has VC
dimension 3. The desired additive accuracy is again roughly $1/V$ for
2-d histograms.

For heatmaps the way density is represented as a color is very
important.  It matters whether the colors are continuously varying
from the background color, or there is an abrupt jump from 0 to 1.
Color scales can also be linear or logarithmic.  Linear color scales
are fairly similar to histograms.  Assume that the maximum probability
for any bucket is $p_{\max}$.  Assume a continuous color scale where
there are $C \approx 20$ discernibly distinct colors, so each
corresponds to an interval of length $p_{\max}/C$. A natural goal in
sampling is that our estimation error should be roughly $p_{\max}/4C$
so that each bucket gets almost the same color as in the ideal
heatmap. This can be achieved using $O(C^2/p_{\max}^2)$
samples. $p_{\max}$ can be as small as $1/HV$ where $H$ and $V$ are
the number of horizontal and vertical pixels, though typically it is
much higher.

Heatmaps where the color is on a log-scale are harder for sampling.
Here we need a constant factor multiplicative approximation to each
$p(b)$ rather than an additive approximation. This is harder to
achieve via sampling, since it requires accurate estimates even for
very small probabilities, and is better done by a 1-pass streaming
algorithm.

\subsection{Heavy Hitters}\label{sec:hitters}

In the heavy hitters problem, we are given a table with $N$ rows, each
holding a value from a domain $D$ and a threshold $\alpha \in [0,1]$.
Let $f(d)$ denote the number of occurrences of item $d \in D$ in the table.
Our goal is to find all $d \in D$ that occur with frequency at
least $\alpha N$.
The naive algorithm is to take a sample of size $n$
and return all elements that occur with frequency at least $3\alpha n/4$.

\begin{theorem}
\label{thm:hh}
  Let $n > \log(1/\alpha\delta)/\alpha^2$. With probability $1 -
  \delta$, the algorithm above returns all $i$ such that $f(i) \geq
  \alpha N$ and does not return any elements where $f(i) \leq \alpha N/4$.
\end{theorem}
\begin{Proof}
We say that an element is heavy if it occurs more than $\alpha N$
times, and light if it occurs fewer than $\alpha N/2$ times.
The Chernoff-Hoeffding bound \cite[Lemma 4.5]{MLbook} implies that the
probability that a heavy element occurs in the sample less than $3\alpha
n/4$ times is bounded by $\exp(-\alpha^2n/16)$. A similar bound holds
for light elements. To finish the argument, we wish to use
the union bound. There are at most $1/\alpha$ heavy elements. There
can be many ($\Omega(N)$) light elements, but we can group them
together into groups so that each group (except one) occurs with frequency in the
range $[\alpha N/4, \alpha N/2]$, and claim that no group occurs too
frequently in our sample. This of course implies that elements in that
group are also not too frequent. Now we apply the union bound, across
at most $5/\alpha$ groups of light elements and $1/\alpha$ heavy
elements. Overall, we want
\[ \frac{6}{\alpha} \exp(-\alpha^2 n/16) \leq \delta \]
which is satisfied for our choice of $n$.
\end{Proof}

\fi
\end{document}